\documentclass{aa} 
\usepackage{graphicx}
\usepackage{txfonts}
\def\etal{{et al. \rm}}
\def\bc{$\beta$ Cephei}

\def\bcma{\object{$\beta$ CMa}}
\def\xcma{\object{$\xi^1$ CMa}}
\def\neri{\object{$\nu$ Eri}}
\def\dcet{\object{$\delta$ Cet}}
\def\voph{\object{V2052 Oph}}
\def\vcen{\object{V836 Cen}}
\def\12lac{\object{12 Lac}}
\def\gpeg{\object{$\gamma$ Peg}}
\def\bcep{\object{$\beta$ Cep}}
\begin{document}
   \title{Abundance analysis of prime B-type targets for asteroseismology}
   \subtitle{I. Nitrogen excess in slowly-rotating \bc \ stars
   \thanks{Table~\ref{tab_ews} is only available in electronic form at the CDS via anonymous ftp to {\tt cdsarc.u-strasbg.fr (130.79.128.5)} or via {\tt http://cdsweb.u-strasbg.fr/cgi-bin/qcat?J/A+A/???/???}}
}
   \titlerunning{Abundance analysis of prime B-type targets for asteroseismology. I}

   \author{T. Morel
          \inst{1}
          \fnmsep\thanks{European Space Agency (ESA) postdoctoral external fellow.}
          \and
          K. Butler
          \inst{2}
          \and
          C. Aerts
          \inst{1,3}
          \and
          C. Neiner
          \inst{1,4}
          \and
          M. Briquet
          \inst{1}
          }

   \offprints{T. Morel, \email{thierry@ster.kuleuven.be}.}

   \institute{Katholieke Universiteit Leuven, Departement Natuurkunde en Sterrenkunde, Instituut voor Sterrenkunde, Celestijnenlaan 200B, B-3001 Leuven, Belgium
         \and
         Universit\"{a}ts-Sternwarte M\"{u}nchen, Scheinerstrasse 1, D-81679 M\"{u}nchen, Germany
         \and
         Department of Astrophysics, University of Nijmegen, PO Box 9010, 6500 GL Nijmegen, The Netherlands
         \and
         GEPI, UMR 8111 du CNRS, Observatoire de Paris-Meudon, 5 place Jules Janssen, 92195 Meudon Cedex, France
}

   \date{Received 9 March 2006 / Accepted 8 July 2006}

\abstract
{Seismic modelling of the \bc \ stars promises major advances in our understanding of the physics of early B-type stars on (or close to) the main sequence. However, a precise knowledge of their physical parameters and metallicity is a prerequisite for correct mode identification and inferences regarding their internal structure. }
{Here we present the results of a detailed NLTE abundance study of nine prime targets for theoretical modelling: \gpeg, \dcet, \neri, \bcma, \xcma, \vcen, \voph, \bcep \ and DD (12) Lac (hereafter \12lac). The following chemical elements are considered: He, C, N, O, Mg, Al, Si, S and Fe. }
{Our curve-of-growth abundance analysis is based on a large number of time-resolved, high-resolution optical spectra covering in most cases the entire oscillation cycle of the stars. }
{Nitrogen is found to be enhanced by up to 0.6 dex in four stars, three of which have severe constraints on their equatorial rotational velocity, $\Omega R$, from seismic or line-profile variation studies: \bcep \ ($\Omega R$$\sim$26 km s$^{-1}$), \voph \ ($\Omega R$$\sim$56 km s$^{-1}$), \dcet \ ($\Omega R$ $<$ 28 km s$^{-1}$) and \xcma \ ($\Omega R\sin i$ $\la$ 10 km s$^{-1}$). The existence of core-processed material at the surface of such largely unevolved, slowly-rotating objects is not predicted by current evolutionary models including rotation. We draw attention to the fact that three stars in this subsample have a detected magnetic field and briefly discuss recent theoretical work pointing to the occurrence of diffusion effects in \bc \ stars possibly capable of altering the nitrogen surface abundance. On the other hand, the abundances of all the other chemical elements considered are, within the errors, indistinguishable from the values found for OB dwarfs in the solar neighbourhood. Despite the mild nitrogen excess observed in some objects, we thus find no evidence for a significantly higher photospheric metal content in the studied \bc \ stars compared to non-pulsating B-type stars of similar characteristics.} 
{}

   \maketitle
%

\section{Introduction} \label{sect_intro}
The class of the \bc \ pulsating variables is defined as B0--B3 V--III stars excited by both (low-order) pressure and gravity modes. The concommitant existence of these two types of modes, along with their relatively simple internal structure, offers the prospect of probing their deep interiors via detailed seismic modelling. This will allow stringent constraints on some fundamental parameters of early B-type dwarfs and subgiants, such as the extent of convective core overshooting or the rotation law in the radiative envelope, to be set in the near future. Although forthcoming or space missions already in operation (e.g. {\em MOST}, {\em COROT}) are expected to lead to dramatic advances in this field, very intensive ground-based observations have already demonstrated the potential of such techniques (Dupret \etal \cite{dupret}). 

The unstable modes in \bc \ stars are driven by the $\kappa$-mechanism and arise from an opacity bump at $T$ $\sim$ 2 $\times$ 10$^5$ K in their interior. It immediately follows that the incidence of pulsations is predicted to be largely controlled by the metal content, with the instability domains in the Hertzprung-Russell (HR) diagram shrinking or even completely vanishing in the low metallicity regimes (Pamyatnykh \cite{pamyatnykh99}; Deng \& Xiong \cite{deng_xiong}). However, recent pieces of evidence suggest that our understanding of the excitation mechanisms in these stars is likely to be still incomplete. First, the number of \bc \ candidates in the LMC is much higher than theoretically predicted (Ko\l aczkowski \etal \cite{kolaczkowski}). Second, current pulsation models have increasing difficulties in accounting for all the frequencies detected in some key objects as more and more intensive observational campaigns are undertaken (e.g. \12lac: Handler \etal \cite{handler}; \neri: Pamyatnykh \etal \cite{pamyatnykh04}; Ausseloos \etal \cite{ausseloos04}). Non-standard stellar models including gravitational settling and radiative levitation might need to be invoked in such cases, as first suggested by Pamyatnykh \etal (\cite{pamyatnykh04}). 

These inconsistencies between theory and observations will be better appraised when the physical parameters (especially $T_{\rm eff}$) and chemical composition (i.e. stellar opacities) of these objects are precisely known. Noteworthy attempts to determine the effective temperature and global metallicity of several \bc \ stars from {\em International Ultraviolet Explorer (IUE)} data have recently been presented (Niemczura \& Daszy\'nska-Daszkiewicz \cite{niemczura_daszynska}), but more robust estimates are likely to be obtained from a detailed analysis of high-quality optical data. Furthermore, a knowledge of the abundances of the individual chemical species is preferable in the context of theoretical modelling, as early-type stars may present significant departures from a scaled solar mixture. In particular, rotationally-induced mixing can lead to the dredge up of some core-processed CNO material in fast-rotating, main-sequence OB stars (e.g. Meynet \& Maeder \cite{meynet_maeder00}; Proffitt \& Quigley \cite{proffitt_quigley}). Such deviations from the standard solar abundance pattern might be routinely incorporated in the near future in any evolution and pulsation codes, and their impact on the pulsational properties explored (see Ausseloos \cite{ausseloos05}).

Notwithstanding the ease with which high-quality spectroscopic data can be gathered for such bright objects, surprisingly little attention has been paid in recent years to derive the stellar parameters and abundances of \bc \ stars from high-resolution optical spectra, with the only dedicated studies going back to the early 70s (Watson \cite{watson71}, \cite{watson72}; Peters \cite{peters}). It is the goal of this project to remedy this situation by taking advantage of the dramatic improvement in data quality, atmospheric models and non-LTE (NLTE) modelling techniques. Here we present a pilot study of nine archetypical \bc \ stars, describing in detail the methodology used to derive the atmospheric parameters and elemental abundances in a self-consistent way. The next papers in this series will be devoted to {a sample of Slowly Pulsating B stars (SPBs) and to the OB-type stars within reach of the asteroseismology programme of the {\em COROT} mission (Baglin \cite{baglin}). 

\section{Observational material} \label{sect_obs}
Most of the selected stars have been the subject of very intensive (multisite) campaigns over the last few years, both in photometric and spectroscopic modes. As such, they are amongst the \bc \ stars (and indeed early-type stars) with the highest number of pulsation modes identified and are thus ideally suited for further theoretical modelling (e.g. \neri: Ausseloos \etal \cite{ausseloos04}; \vcen: Aerts \etal \cite{aerts03}). Table~\ref{tab_pulsations} summarizes the basic pulsation properties of our targets. Most of the stars with a dominant radial mode do not possess large-amplitude non-radial modes, except \neri. On the other hand, the radial mode is excited in the three stars with a dominant non-radial mode, but with a lower amplitude.

\begin{table}
\centering
\caption{Basic pulsation properties of our targets. $N$($\nu$): number of independent pulsation frequencies detected. Data from stated papers and references therein.}
\label{tab_pulsations}
\begin{tabular}{llcl}
\hline
Star   & Dominant pulsation   & $N$($\nu$)  & Most recent\\
       & mode                 &             & reference   \\\hline
\gpeg  & Radial               &   4         & Chapellier \etal (\cite{chapellier})\\
\dcet  & Radial               &   4         & Aerts \etal (\cite{aerts06})\\
\neri  & Radial               &  14         & Jerzykiewicz \etal (\cite{jerzykiewicz})\\
\bcma  & Non-radial           &   3         & Desmet \etal (\cite{desmet})\\
\xcma  & Radial               &   1         & Saesen \etal (\cite{saesen})\\
\vcen  & Non-radial           &   6         & Aerts \etal (\cite{aerts04b})\\
\voph  & Radial               &   2         & Neiner \etal (\cite{neiner})\\
\bcep  & Radial               &   5         & Telting \etal (\cite{telting97})\\
\12lac & Non-radial           &  11         & Handler \etal (\cite{handler})\\
\hline 
\end{tabular}\\
\end{table}

A description of the spectroscopic data used is given in Table~\ref{tab_obs}. Further details on the data acquisition and reduction procedures can be found in Aerts \etal (\cite{aerts04a}; \neri), Mazumdar \etal (\cite{mazumdar}; \bcma), Saesen \etal (\cite{saesen}; \xcma), Aerts \etal (\cite{aerts04b}; \vcen), Neiner et al., in prep. (\voph) and Desmet et al., in prep. (\12lac). The data for \dcet \ were obtained at ESO during the 2002 July 17--25 period and were reduced with a dedicated reduction software for the CORALIE spectrograph (see Baranne \etal \cite{baranne}). The spectra of \gpeg \ and \bcep \ have been retrieved from the ELODIE archives (see Moultaka \etal \cite{moultaka}). Aerts \etal (\cite{aerts04b}) presented a preliminary spectral synthesis analysis of \vcen \ based on a spectrum reduced with the FEROS reduction pipeline. The raw data have been completely re-reduced here with standard {\tt IRAF}\footnote{{\tt IRAF} is distributed by the National Optical Astronomy Observatories, operated by the Association of Universities for Research in Astronomy, Inc., under cooperative agreement with the National Science Foundation.} tasks, resulting in a much higher data quality. 

Our analysis for each star is based on a mean spectrum generally created by co-adding a large number of individual exposures (up to 579; Table~\ref{tab_obs}). These spectra were put in the laboratory rest frame prior to forming the weighted (by the S/N ratio) average of the whole time series, as large radial velocity variations are often observed. The orders of the mean spectrum were then merged using {\tt IRAF} tasks. Sections of the merged spectrum were subsequently continuum normalized by fitting the line-free regions with low-order cubic spline polynomials. Each individual echelle order was separately normalized for the data acquired with the CS2 spectrograph at McDonald observatory, as the merging procedure proved unsatisfactory in that case. 

\begin{table*}
\centering
\caption{Basic description of the spectroscopic data. Spectral types from the SIMBAD database (revised values are given in Table~\ref{tab_parameters}). $\Delta \lambda$: spectral range covered in \AA, $R$: mean resolving power of the spectrograph, $N$: number of exposures combined, $S/N$: typical signal-to-noise ratio at about 4500 \AA \ in the {\em individual} exposures (estimated from photon statistics).}
\label{tab_obs}
\scriptsize
\begin{tabular}{lcccccccccc} \hline\hline
                 & \gpeg           & \dcet             & \neri             & \bcma             & \multicolumn{2}{c}{\xcma$^a$}         & \vcen              & \voph              & \bcep              & \12lac            \\
                 &                 &                   &                   &                   & Max EWs & Min EWs                     &                    &                    &                    &                   \\\hline
HD number       & \object{HD 886} & \object{HD 16582} & \object{HD 29248} & \object{HD 44743} & \multicolumn{2}{c}{\object{HD 46328}} & \object{HD 129\,929} & \object{HD 163\,472} & \object{HD 205\,021} & \object{HD 214\,993} \\
Spectral type    & B2 IV           & B2 IV             & B2 III            & B1 II/III         & \multicolumn{2}{c}{B1 III}            & B3 V               & B2 IV--V           & B2 IIIevar         & B2 III             \\
Telescope        & 1.9-m OHP       & 1.2-m Euler         & 1.2-m Euler         & 1.2-m Euler         & \multicolumn{2}{c}{1.2-m Euler}         & 2.2-m Euler          & 2.7-m McDonald     & 1.9-m OHP          & 2.7-m McDonald     \\
Instrument       & ELODIE          & CORALIE           & CORALIE           & CORALIE           & \multicolumn{2}{c}{CORALIE}           & FEROS              & CS2                & ELODIE             & CS2                \\
$\Delta \lambda$ & 3895--6820      & 3875--6820        & 3875--6820        & 3875--6820        & \multicolumn{2}{c}{3875--6820}        & 3555--9215         & 3630--10\,275      & 3895--6820         & 3630--10\,275      \\
$R$              & 50\,000         & 50\,000           & 50\,000           & 50\,000           & \multicolumn{2}{c}{50\,000}           & 48\,000            & 60\,000            & 50\,000            & 60\,000           \\
$N$              & 47              & 4                 & 579               & 449               & 38               & 41                 & 1                  & 105                & 28                 & 31                 \\
$S/N$            & 215             & 150               & 250               & 250               & 180              & 200                & 150                & 230                & 215                & 285                 \\
\hline
\end{tabular}
\begin{flushleft}
$^a$ Min and Max EWs correspond to minimum and maximum EWs of the  \ion{Si}{iii} lines, respectively.
\end{flushleft}
\end{table*}

The very high signal-to-noise ratio often attained for the combined spectrum enables us to confidently measure weak diagnostic lines. Figure~\ref{fig_iron} illustrates this point for some \ion{Fe}{iii} features and allows one to get a sense of the overall data quality. More importantly, combining such a large number of time-resolved spectra also allows us to minimize the impact of the strong line-profile variations arising from pulsations (see Aerts \etal \cite{aerts94} for some examples) and the related changes in the equivalent widths (EWs), which are, for instance, typically of the order of 3--4\% for the \ion{Si}{iii} lines in \bc \ stars (De Ridder \etal \cite{deridder02}). All the fundamental quantities we derive (e.g. $T_{\rm eff}$) can thus be regarded in most cases as representative of the mean values averaged over the pulsation cycles. This is one of the most important aspects of the present work with respect to previous studies based on snapshot spectra (e.g. Gies \& Lambert \cite{gies_lambert}; hereafter GL). In the case of \xcma, the excellent time sampling and relatively simple nature of the pulsations (only a large-amplitude radial mode is excited) prompted us to examine the phase-related changes in the atmospheric parameters. To this end, we separately analyzed two mean spectra created for two phase intervals of width $\Delta \phi$=0.1 corresponding to minimum and maximum EWs of the  \ion{Si}{iii} lines (see Saesen \etal \cite{saesen}). 

\begin{figure}
\centering
\includegraphics[width=8.7cm]{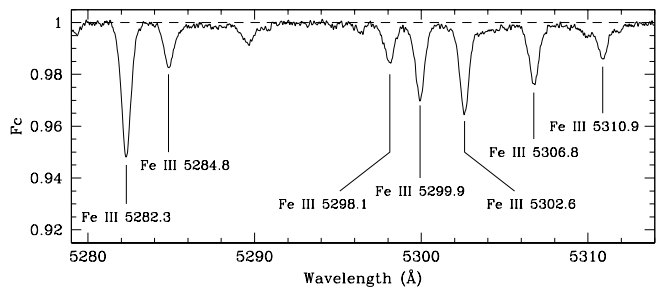}
\caption{Mean, normalized spectrum of \bcma \ in the spectral range 5279--5314 \AA. Some weak \ion{Fe}{iii} lines are indicated.}
\label{fig_iron}
\end{figure}

\section{Method of analysis} \label{sect_analysis}
\subsection{Model atmosphere calculations and spectral synthesis} \label{sect_models}
We made use of the latest versions of the NLTE line formation codes DETAIL and SURFACE originally developed by Butler (\cite{butler}) and Giddings (\cite{giddings}), along with plane-parallel, fully line-blanketed LTE Kurucz atmospheric models (Kurucz \cite{kurucz93}). Grids with solar metallicity were used. Slight deviations from the solar mixture for some metals (e.g. N; see below) do not have an appreciable effect on the atmospheric structure. We adopted models with He/H=0.089 by number in all cases, in accordance with the derived helium abundances. The only exception was \voph, as it displays some evidence for a helium enrichment (see Sect.~\ref{sect_results} and Neiner \etal \cite{neiner}). 

We used Kurucz Opacity Distribution Functions (ODFs) with a fine frequency mesh (1212 points). To limit computational time, we adopted a constant microturbulent velocity, $\xi_{\rm mod}$=8 km s$^{-1}$, in the statistical equilibrium calculations with DETAIL. This value is in most cases consistent with the microturbulence derived from the line analysis. Test calculations for \bcma \ show that the exact choice of this parameter has little impact on the final results, with the mean abundance differences being less than 0.05 dex for all elements when adopting $\xi_{\rm mod}$=2 or 8 km s$^{-1}$. 

The following chemical elements are considered: He, C, N, O, Mg, Al, Si, S and Fe. The number of levels for the various ionic species is provided in Table~\ref{tab_levels}, along with the original literature sources where basic information on the model atoms can be found. Note that the atomic data used  (e.g. photoionization cross sections) have been regularly updated following advances in theoretical calculations. The \ion{Fe}{iii} model atom will be described in detail in a forthcoming paper (Butler et al., in prep.).  Here we note that the bulk of the atomic data, energy levels, oscillator strengths and damping parameters are taken from Kurucz (\cite{kurucz93}). While the \ion{Fe}{ii} and \ion{Fe}{iv} atoms are rudimentary, the \ion{Fe}{iii} model includes all terms calculated to lie below the ionization threshold.

The line atomic data (e.g. oscillator strengths, radiative damping constants) are taken from the NIST\footnote{Available online at: {\tt http://physlab2.nist.gov/PhysRefData/\linebreak ASD/index.html}.} or VALD (Kupka \etal \cite{kupka}) databases. The NLTE populations are computed assuming LS coupling. Our line list includes most of the diagnostic lines generally used in abundance analyses of early B-type stars, but great care has been taken to only retain unblended features for the relevant temperature ranges.\footnote{We made extensive use of the spectral atlases for main-sequence B stars available at: {\tt http://www.lsw.uni-heidelberg.de/\linebreak cgi-bin/websynspec.cgi} (see Gummersbach \etal \cite{gummersbach}).} Our final line list is made up of about 180 spectral lines. 

\begin{table}
\centering
\caption{Number of levels for the various ions having their populations explicitely treated in NLTE with DETAIL. The ground states of \ion{C}{iv}, \ion{N}{iv}, \ion{O}{iv},  \ion{Mg}{iii}, \ion{Al}{iv}, \ion{Si}{v}, \ion{S}{iv} and \ion{Fe}{v} are also considered.}
\label{tab_levels}
\begin{tabular}{lcl}
\hline
Ion     & Number     & Reference\\
        & of levels  &\\\hline
\ion{H}{i}    & 10  & Husfeld \etal (\cite{husfeld})               \\
\ion{He}{i}   & 27  & Husfeld \etal (\cite{husfeld})                  \\
\ion{He}{ii}  & 14  &                   \\
\ion{C}{ii}   & 54  & Eber \& Butler (\cite{eber_butler})\\
\ion{C}{iii}  & 3   &\\
\ion{N}{i}    & 3   & Becker \& Butler (\cite{becker_butler89})\\
\ion{N}{ii}   & 50  &\\
\ion{N}{iii}  & 5   &\\
\ion{O}{i}    & 3   & Becker \& Butler (\cite{becker_butler88})\\
\ion{O}{ii}   & 52  &\\
\ion{O}{iii}  & 3   &\\
\ion{Mg}{i}   & 88  & Przybilla \etal (\cite{przybilla01})\\
\ion{Mg}{ii}  & 37  &\\
\ion{Al}{iii} & 12  & Dufton \etal (\cite{dufton86})\\
\ion{Si}{ii}  & 34  & Trundle \etal (\cite{trundle}), see also text\\
\ion{Si}{ii}  & 28  &\\
\ion{Si}{iv}  & 18  &\\
\ion{S}{ii}   & 78  & Vrancken \etal (\cite{vrancken96})\\
\ion{S}{iii}  & 21  &\\
\ion{Fe}{ii}  & 8   & Butler et al., in prep.\\
\ion{Fe}{iii} & 264 & \\
\ion{Fe}{iv}  & 16  & \\
\hline 
\end{tabular}\\
\end{table}

As will be shown in Sect.~\ref{sect_results}, the programme stars span a restricted range in $T_{\rm eff}$ and $\log g$. This implies that the departures from LTE should be roughly uniform among our sample and the relative abundances fairly insensitive to the treatment of NLTE line formation. Extremely few NLTE iron abundance studies of early B-type stars have been presented in the literature (to our knowledge, the only one being the analysis of the B1.5 dwarf \object{HD 35299} by Vrancken \cite{vrancken97}). The magnitude of the NLTE corrections affecting the \ion{Fe}{iii} lines, $\Delta \epsilon$=($\log\epsilon)_{\rm NLTE}$--($\log\epsilon)_{\rm LTE}$, hence deserves further discussion. The departures from LTE are found to negligible for the coolest objects in our sample ($\Delta \epsilon$ $\la$ +0.05 dex), but to significantly increase with $T_{\rm eff}$ (e.g. $\Delta \epsilon$=+0.21 dex for \bcep \ and +0.33 dex for \xcma, respectively). Previous LTE values in the literature derived from \ion{Fe}{iii} lines for B1 dwarfs or earlier subtypes can thus be regarded as strict lower limits. For the other elements, an illustrative LTE abundance analysis carried out on \bcma \ suggests systematically negative mean NLTE corrections never exceeding 0.17 dex (see Table~\ref{tab_nlte}). Such modest departures from LTE for early B-type dwarfs or subgiants explains the reasonable agreement with past LTE abundance determinations in the literature (e.g. Martin \cite{martin} and Rolleston \etal \cite{rolleston03} in the case of \gpeg \ and \xcma, respectively); large discrepancies can generally be ascribed to other factors, such as differences in the adopted atmospheric parameters or atomic data.

\begin{table}
\centering
\caption{NLTE corrections in the case of \bcma. The number of lines considered is given in brackets. Note that the listed corrections for some ions are based on very few lines and may not be representative.}
\label{tab_nlte}
\begin{tabular}{llc} \hline\hline
Element & Ion                 & $\Delta \epsilon$=($\log\epsilon)_{\rm NLTE}$--($\log\epsilon)_{\rm LTE}$\\\hline
C  & \ion{C}{ii} (8)                             & --0.07\\ 
   & \ion{C}{iii} (1)                            &   0.00\\ 
   & \ion{C}{ii}+\ion{C}{iii} (9)                & --0.06\\
N  & \ion{N}{ii} (25)                            & --0.11\\ 
   & \ion{N}{iii} (1)                            &   0.00\\ 
   & \ion{N}{ii}+\ion{N}{iii} (26)               & --0.11\\
O  & \ion{O}{ii}  (30)                           & --0.14\\
Mg & \ion{Mg}{ii} (2)                            & --0.10\\ 
Al & \ion{Al}{iii} (4)                           &   0.00\\ 
Si & \ion{Si}{ii} (2)                            & +0.03\\ 
   & \ion{Si}{iii} (6)                           & --0.26\\ 
   & \ion{Si}{iv} (1)                            & --0.06\\ 
   & \ion{Si}{ii}+\ion{Si}{iii}+\ion{Si}{iv} (9) & --0.17\\
S  & \ion{S}{ii} (1)                             & +0.10\\ 
   & \ion{S}{iii} (1)                            & --0.12\\ 
   & \ion{S}{ii}+\ion{S}{iii} (2)                & --0.01\\ 
Fe & \ion{Fe}{iii} (25)                          & +0.09\\ 
\hline
\end{tabular}
\end{table}

\subsection{Estimate of atmospheric parameters} \label{sect_parameters}
The stellar physical parameters were derived in a standard way, whereby $T_{\rm eff}$ and $\log g$ are self-consistently determined using an iterative scheme. Few iterations are needed before convergence is attained when sensible starting values can easily be estimated a priori, as is the case here. The effective temperature was estimated from the \ion{Si}{ii/iii/iv} ionization equilibrium. Unfortunately, the spectral lines of these three ions are only measurable in \bcma \ and \bcep. For the other stars, we had to solely rely on the \ion{Si}{ii/iii} or \ion{Si}{iii/iv} lines. This has motivated us to replace the 12-level \ion{Si}{ii} model ion initially implemented in DETAIL (Becker \& Butler \cite{becker_butler90}) by a more complete atomic model with 34 levels developed by D.~J.~Lennon. This is expected to allow for a more realistic treatment of the ionization balance and leads to a slightly cooler temperature scale (see Trundle \& Lennon \cite{trundle_lennon}). Figure~\ref{fig_ratio} provides various examples of calibrations between the Si line ratios and $T_{\rm eff}$. We used solar Si abundances (as appropriate for our stars; see below) and atmospheric models with a $\xi_{\rm mod}$ value matching as closely as possible the microturbulence estimated from the line analysis. Several line ratios (up to 12 in total) were used for the temperature determination. Not all of them are completely independent (only seven transitions are involved), but consistent results were obtained in all cases. The dispersion between the obtained values was taken as representative of the temperature uncertainty. This proved to be comparable from star to star and of the order of 1000 K. In cases where the abundances of C, N and S are based on lines of two adjacent ionization stages, no convincing evidence for systematic discrepancies was found. As a result, all abundances yielded by the individual lines were averaged irrespective of their ionization stage. This gives credence to the reliability of our temperature scale, although the scope of such a comparison is limited by the coarse treatment of the minor ionic species \ion{C}{iii} and \ion{N}{iii} in our model atoms (Table~\ref{tab_levels}). We will comment further on this point in Sect.~\ref{sect_results_parameters} by comparing our $T_{\rm eff}$ values with literature data. 

\begin{figure}
\centering
\includegraphics[width=8.5cm]{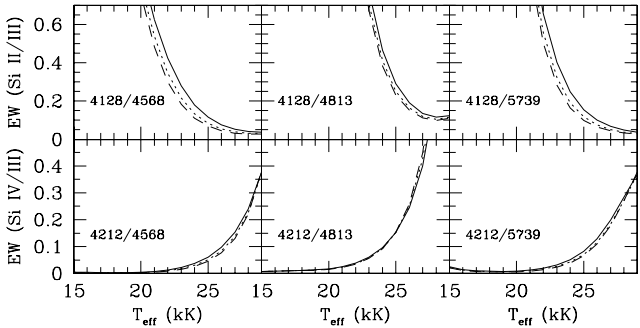}
\caption{Examples of calibrations between various Si line ratios and the effective temperature, as a function of $\xi$. The dependence on the microturbulence must in certain cases be taken into account, as the lines used may be of very different strength. Solid line: $\xi$=5 km s$^{-1}$; dotted line: $\xi$=10 km s$^{-1}$; dashed line: $\xi$=15 km s$^{-1}$. In this case we adopted $\log g$=3.5 [cgs], $\log \epsilon$(Si)=7.20 dex and $\xi_{\rm mod}$=8 km s$^{-1}$.}
\label{fig_ratio}
\end{figure}

The surface gravity was derived by fitting the collisionally-broadened wings of the Balmer lines. Line broadening mechanisms such as rotation or microturbulence only affect the line cores and hence no prior knowledge of these quantities is needed. The H$\beta$ line was not used in \bcep, as the H$\alpha$ profile is clearly filled-in by a variable emission component. H$\beta$ may thus slightly be affected by emission as well. All of H$\epsilon$, H$\delta$, H$\gamma$ and H$\beta$ were used for the other stars, as the lower series members are unlikely to be substantially affected by circumstellar/wind emission. Indeed, they systematically yielded consistent results, as can be seen in Fig.~\ref{fig_balmer} in the case of \bcma. Typical uncertainties are of the order of 0.15--0.20 dex and reflect both the dispersion of the values obtained using the various lines, errors arising from the uncertainty on the $T_{\rm eff}$ calibration and imperfections in the merging of the echelle orders. As mentioned in Sect.~\ref{sect_obs}, this step in the reduction procedure was not possible for the McDonald data. Therefore, the profile fitting of the Balmer lines in \voph \ was performed using spectra obtained with the Coud\'e Echelle spectrograph mounted on the 2-m telescope of the Tautenburg observatory, Germany (see Neiner et al., in prep.). For \12lac, we made use of a snapshot spectrum extracted from the ELODIE archives.

\begin{figure}
\centering
\includegraphics[width=8.5cm]{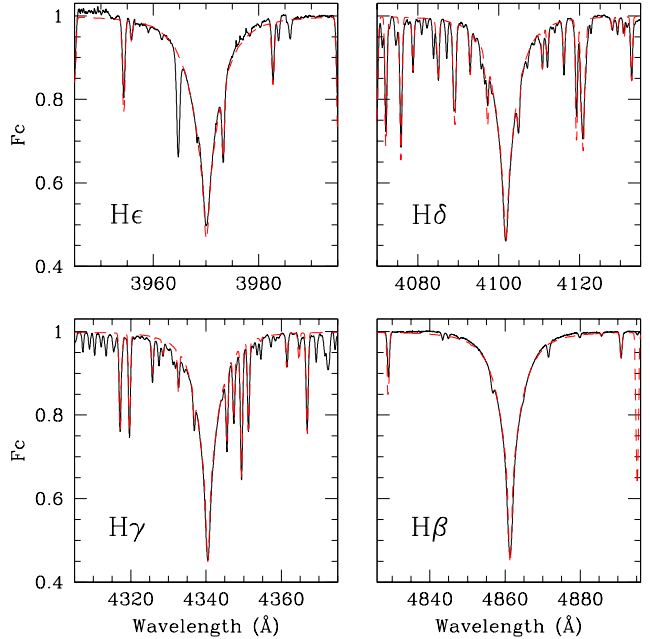}
\caption{Synthetic ({\em dotted line}; $T_{\rm eff}$=24\,000 K and $\log g$=3.5 [cgs]) and observed ({\em solid line}) spectra of \bcma \ for the regions encompassing the Balmer lines.}
\label{fig_balmer}
\end{figure}

The microturbulence, $\xi$, was estimated by requiring the abundances yielded by the individual \ion{O}{ii} lines to be independent of the line strength. To estimate the uncertainties, we varied this parameter until the slope of the $\log \epsilon$(O)-$\log$($EW/\lambda$) relation differs from zero at the 3$\sigma$ level. The oxygen features are the strongest metal lines in our spectra and are thus best suited for this purpose. However, the microturbulent velocities obtained from the \ion{N}{ii} lines are much lower than those obtained from the \ion{O}{ii} features for the most massive objects (\bcma, \xcma, \bcep \ and \12lac) and are in better agreement with the values inferred for B-type stars in general using other methods (e.g. Fitzpatrick \& Massa \cite{fitzpatrick_massa}). Although extensively discussed in the literature, the possible causes of these inconsistencies remain unclear (e.g. Trundle \etal 2004; see also Sect.~\ref{sect_wind}). We provide in Table~\ref{tab_micro} some illustrative examples of the abundance differences resulting from the exact choice of the species used to determine the microturbulence. Although we note that adopting the low values suggested by the \ion{N}{ii} lines would lead to unreasonably high abundances for several strong lines of other elements, we cannot rule out a systematic, albeit slight ($\Delta\log \epsilon$ $\lesssim$ 0.2 dex), underestimate of the stellar abundances in some objects.

\begin{table}
\centering
\caption{Abundance differences when adopting the microturbulent velocities yielded by the \ion{N}{ii} lines ($\xi$=2 and 3 km s$^{-1}$ for \xcma \ and \bcep, respectively) instead of the default values derived from the \ion{O}{ii} features ($\xi$=6 km s$^{-1}$ in both cases).}
\label{tab_micro}
\begin{tabular}{lcc} \hline\hline
                          & \xcma                        & \bcep \\
$\Delta \xi$              & from 6 to 2 km s$^{-1}$ & from 6 to 3 km s$^{-1}$\\\hline
$\Delta$He/H              & +0.014 & +0.005\\
$\Delta\log \epsilon$(C)  & +0.02  & +0.01\\ 
$\Delta\log \epsilon$(N)  & +0.08  & +0.05\\ 
$\Delta\log \epsilon$(O)  & +0.20  & +0.13\\ 
$\Delta\log \epsilon$(Mg) & +0.18  & +0.14\\ 
$\Delta\log \epsilon$(Al) & +0.10  & +0.06\\ 
$\Delta\log \epsilon$(Si) & +0.09  & +0.17\\ 
$\Delta\log \epsilon$(S)  & +0.04  & +0.02\\ 
$\Delta\log \epsilon$(Fe) & +0.06  & +0.04\\ 
\hline
\end{tabular}
\end{table}

\subsection{Estimate of chemical abundances} \label{sect_abundances}
The abundances are derived once the atmospheric parameters mentioned above are known by matching the measured EWs of the selected lines (tabulated in Table~\ref{tab_ews}) and the values measured in the synthetic spectra. Such an EW-based abundance analysis is made possible by the fact that all our stars are relatively slow rotators, as will be shown below. For consistency, direct integration was used in both cases. In a few instances, multicomponent fitting using a Voigt profile was performed with {\tt IRAF} tasks in the case of well-separated blends. 

Five sources of error on the final abundances were considered. Apart from the line-to-line scatter\footnote{The Mg abundance was often derived from a single line. In this case, we fixed $\sigma_{\rm int}$ to 0.15 dex.} ($\sigma_{\rm int}$), we first calculated the errors arising from the uncertainties on the atmospheric parameters ($\sigma_{T_{\rm eff}}$, $\sigma_{\log g}$ and $\sigma_{\xi}$). They were derived by computing the abundances using models with atmospheric parameters deviating from the final values by the relevant uncertainties. We also took into account the fact that the determinations of $T_{\rm eff}$ and $\log g$ are strongly coupled ($\sigma_{T_{\rm eff}/\log g}$). The other covariance terms do not significantly contribute to the total error budget and can be neglected. Finally, we quadratically add these errors to obtain the total uncertainty, $\sigma_T$. Table~\ref{tab_errors} illustrates these calculations in the case of \bcma. It can be seen that the accuracy of the abundance determination varies significantly depending upon the element considered, being typically of the order of 0.1 dex for C, but much larger for S and Si, for instance ($\sim$0.2--0.3 dex). This primarily reflects the sensitivity to changes in the physical parameters, limitations in the model atoms, as well as inaccuracies in the atomic data. 
\begin{table}
\centering
\caption{Calculation of the error budget in the case of \bcma. $\sigma_{\rm int}$: line-to-line scatter; $\sigma_{T_{\rm eff}}$: variation of the abundances for $\Delta T_{\rm eff}$=+1000 K; $\sigma_{\log g}$: as before, but for $\Delta \log g$=+0.15 dex; $\sigma_{\xi}$: as before, but for $\Delta \xi$=+3 km s$^{-1}$; $\sigma_{T_{\rm eff}/\log g}$: as before, but for $\Delta T_{\rm eff}$=+1000 K and $\Delta \log g$=+0.15 dex; $\sigma_T$: total uncertainty.}
\label{tab_errors}
\begin{tabular}{lcccccc} \hline\hline
                           & $\sigma_{\rm int}$ & $\sigma_{T_{\rm eff}}$ & $\sigma_{\log g}$ & $\sigma_{\xi}$ & $\sigma_{T_{\rm eff}/\log g}$  & $\sigma_T$ \\\hline
$\Delta$He/H              & 0.014 & 0.010 & 0.007 & 0.020 & 0.003 & 0.028\\
$\Delta\log \epsilon$(C)  & 0.037 & 0.074 & 0.001 & 0.019 & 0.059 & 0.103\\ 
$\Delta\log \epsilon$(N)  & 0.121 & 0.046 & 0.008 & 0.026 & 0.042 & 0.139\\ 
$\Delta\log \epsilon$(O)  & 0.088 & 0.078 & 0.040 & 0.109 & 0.053 & 0.174\\ 
$\Delta\log \epsilon$(Mg) & 0.156 & 0.060 & 0.015 & 0.050 & 0.050 & 0.182\\ 
$\Delta\log \epsilon$(Al) & 0.096 & 0.093 & 0.022 & 0.032 & 0.060 & 0.152\\ 
$\Delta\log \epsilon$(Si) & 0.206 & 0.036 & 0.009 & 0.086 & 0.028 & 0.228\\ 
$\Delta\log \epsilon$(S)  & 0.212 & 0.100 & 0.035 & 0.015 & 0.060 & 0.245\\ 
$\Delta\log \epsilon$(Fe) & 0.178 & 0.033 & 0.031 & 0.015 & 0.046 & 0.190\\ 
\hline
\end{tabular}
\end{table}

\begin{table*}
\centering
\caption{Physical parameters of the programme stars. The revised MK spectral types are based on our $T_{\rm eff}$ estimates and the calibration of Crowther (\cite{crowther}). The 1-$\sigma$ uncertainty on $T_{\rm eff}$ and $\log g$ are 1000 K and 0.15 dex, respectively (except for \vcen \ and \voph: $\Delta \log g$=0.20 dex).}
\hspace*{-1.3cm}
\scriptsize
\label{tab_parameters}
\begin{tabular}{lcccccccccc} \hline\hline
                           & \gpeg           &  \dcet          & \neri           & \bcma           & \multicolumn{2}{c}{\xcma}         & \vcen           & \voph           & \bcep & \12lac\\
                           &                 &                 &                 &                 & Max EWs & Min EWs                 &                 &                 &   &        \\\hline
Revised spectral type & B1.5--B2 V & B1.5--B2 V & B1.5--B2 V & B1.5 V--IV & \multicolumn{2}{c}{B0.5--B1 V--IV} & B1.5 V & B1.5--B2 V & B1 Vevar &  B1.5 V\\
$T_{\rm eff}$ (K)      & 22\,500         & 23\,000         & 23\,500         & 24\,000         & 27\,000   & 28\,000               & 24\,500         & 23\,000         & 26\,000      &  24\,500       \\ 
$\log T_{\rm eff}$ (K)     & 4.352$\pm$0.019 & 4.362$\pm$0.019 & 4.371$\pm$0.019 & 4.380$\pm$0.018 & 4.431$\pm$0.016 & 4.447$\pm$0.016 & 4.389$\pm$0.018 & 4.362$\pm$0.019 & 4.415$\pm$0.017      & 4.389$\pm$0.018       \\ 
$\log g$ (cm s$^{-2}$) & 3.75            & 3.80            & 3.75            & 3.50            & 3.70      & 3.80                  & 3.95            & 4.00             & 3.70      &  3.65       \\ 
$\xi$ (km s$^{-1}$)        & 1$^{+2}_{-1}$   & 1$^{+3}_{-1}$   & 10$\pm$4        & 14$\pm$3        & \multicolumn{2}{c}{6$\pm$2}       & 6$\pm$3         & 1$^{+4}_{-1}$   &  6$\pm$3     &   10$\pm$4      \\
\hline 
$v_T$ (km s$^{-1}$)              & 10      & 14      & 36       & 23       & \multicolumn{2}{c}{10} & 15      & 62       & 29      & 42 \\ 
$\left[\left(\Omega R \sin i\right)^2 + v_{\rm macro}^2\right]^{1/2}$ (km s$^{-1}$) & $\sim$10      & $\sim$13      & $\sim$16       & $\sim$20       & \multicolumn{2}{c}{$\sim$10} &         & $\sim$61       & $\sim$29      & $\sim$37 \\
$v_{\rm puls}$ (km s$^{-1}$)     & negligible & negligible & $\sim$32 & $\sim$11 & \multicolumn{2}{c}{negligible}                &         & negligible & negligible & $\sim$20\\
$\Omega R$ (km s$^{-1}$)         &         &   14 or 28$^a$      & 6$^a$    & 32$^b$   &        &               & 2$^a$   & 56$^c$   & 26$^c$  & 45$^b$\\
$v_{\rm macro}$  (km s$^{-1}$)   &  & ? & $\gtrsim$ 15 & ? &  \multicolumn{2}{c}{} &  & $\gtrsim$ 20 & $\gtrsim$ 10 & ?\\
\hline
\end{tabular}
\begin{flushleft}
$^a$ From seismic studies (Dupret \etal \cite{dupret}; Pamyatnykh \etal \cite{pamyatnykh04}; Aerts \etal \cite{aerts06}).\\ 
$^b$ From modelling of the line-profile variations (the so-called ``moment method''; Aerts \cite{aerts96}, Desmet \etal \cite{desmet}).\\
$^c$ Calculated from the stellar radius derived from evolutionary tracks (Fig.~\ref{fig_hr}) and assuming that the rotational period can be identified with the recurrence timescale of the changes affecting the UV lines: $\cal{P_{\rm rot}}$=12.001\,06 d in \bcep \ (Henrichs \etal \cite{henrichs}) and  3.638\,833 d in \voph \ (Neiner \etal \cite{neiner}).
\end{flushleft}
\end{table*}

A comment is necessary regarding the helium abundance determination in \voph. This star has been classified as He-strong by Neiner \etal (\cite{neiner}) who obtained He/H=0.21 by number based on a detailed NLTE analysis of a number of optical \ion{He}{i} lines. Although our curve-of-growth analysis does not support such a high helium content (see below), contrary to the other stars in our sample we were unable to obtain consistent fits for all the \ion{He}{i} lines considered. In particular, it was impossible to fit the wings of the diffuse lines using a single helium abundance, some lines being well fit with He/H=0.09, others with twice the solar value (particularly the singlets). Interestingly, a similar phenomenon is observed in \bcep. A detailed investigation is beyond the scope of this paper, but this phenomenon may be related to the existence of surface inhomogeneities/vertical stratification of helium in these two magnetic stars. As mentioned in Sect.~\ref{sect_models}, Kurucz models with He/H=0.178 have been used for \voph. This choice has a negligible impact on the atmospheric parameters and abundances ($\Delta$He/H=0.003; $\Delta \log \epsilon$ $\la$ 0.05 dex).

\subsection{Estimate of the amount of line broadening} \label{sect_vsini}
We provide in Table~\ref{tab_parameters} a total line broadening parameter, $v_T$, defined as:
\begin{equation}
\label{equ_1}
v_T^2 = \left(\Omega R \sin i\right)^2 + v_{\rm macro}^2 + v_{\rm puls}^2,
\end{equation}
with $\Omega$ the equatorial angular rotational frequency, $R$ the stellar radius, $i$ the inclination angle, $v_{\rm macro}$ the macroturbulent velocity and $v_{\rm puls}$ the amount of pulsational line broadening averaged over the oscillation cycle. The latter is expected to be significant for non-radial modes, but to be small for radial pulsators (cf. comparison between Tables~\ref{tab_pulsations} and ~\ref{tab_parameters}). The amount of macroturbulent broadening is unknown for our objects which are on, or close to, the main sequence. The $v_T$ values were estimated as a final step of the abundance analysis by fitting the profiles of some isolated \ion{O}{ii} lines with a grid of synthetic spectra with the appropriate microturbulent velocity and convolved with a rotational broadening profile (Gray \cite{gray}), assuming a limb-darkening coefficient, $\epsilon$=0.4 (Claret \cite{claret}). The instrumental profile was estimated from calibration lamps. 

To disentangle the effect of rotational and macroturbulent broadening on the one hand, and of the averaged pulsational line broadening on the other hand, we have selected the spectrum least affected by pulsations in the time series, i.e. with the narrowest and most symmetric line profiles (this is not possible for \vcen, as we only have one observation). These were used to derive the quantity: $\left(\Omega R \sin i\right)^2 + v_{\rm macro}^2$. The amount of averaged pulsational line broadening immediately follows from Eq.~\ref{equ_1}. The true stellar rotation rates, $\Omega R$, are known for most targets (Sect.~\ref{sect_mixing}) and indicate that some macroturbulence does seem to occur in some stars (see Table~\ref{tab_parameters}).

\section{Results and comparison with previous studies} \label{sect_results}
\subsection{Physical parameters}\label{sect_results_parameters}
The atmospheric parameters of the programme stars are presented in Table~\ref{tab_parameters}. We also provide the revised MK spectral types based on our $T_{\rm eff}$ estimates and the spectral type-$T_{\rm eff}$ calibration for Galactic OB dwarfs of Crowther (\cite{crowther}). It is necessary to assess the reliability of our derived $T_{\rm eff}$ and $\log g$ estimates before discussing the chemical properties of our targets. An exhaustive literature search was not attempted. Instead, a comparison was made with atmospheric parameters obtained using three independent methods: (i) $T_{\rm eff}$ primarily derived from Str\" omgren photometry and $\log g$ from profile fitting of H$\gamma$ (GL); (2) both parameters estimated from Walraven colour indices, or from Geneva and Str\" omgren photometry when these data were not available (Heynderickx \etal \cite{heynderickx}); (iii) temperatures based on the continuum fitting of low-resolution {\em IUE} data (Niemczura \& Daszy\'nska-Daszkiewicz \cite{niemczura_daszynska}). As can be seen in Fig.~\ref{fig_teff}, our temperatures (and consequently surface gravities) are systematically lower than the values quoted by GL, but there are reasons to believe that their temperature scale is too hot. In particular, they applied a systematic upward correction to their temperatures in order to match the empirical calibration derived by Code \etal (\cite{code}) from angular diameter and absolute flux measurements. However, more recent and accurate observations (Smalley \& Dworetsky \cite{smalley_dworetsky}) lead to systematically lower values by about 1000 K in this temperature range (see discussion in Lyubimkov \etal \cite{lyubimkov02}). The star \bcma \ was incidentally included in this sample of standard stars and is ideally suited for a direct comparison: $T_{\rm eff}$ was decreased from 25\,180$\pm$1130 (Code \etal \cite{code}) to 24\,020$\pm$1150 K (Smalley \& Dworetsky \cite{smalley_dworetsky}), which is now in better agreement with our results. Additionally, we note that Aufdenberg \etal (\cite{aufdenberg}) obtained absolutely identical values to ours for $T_{\rm eff}$ and $\log g$ from the fitting of the spectral energy distribution from the near-UV to the infrared using a fully line-blanketed, spherical and NLTE atmospheric model. On the other hand, our $T_{\rm eff}$ values tend to be slightly higher than those obtained from colour indices. The photometric calibrations in the Walraven photometric system are based on the grids of lightly line-blanketed model atmospheres of Kurucz (\cite{kurucz79}). Unfortunately, the lack of further details (e.g. zero points) does not allow us to identify the cause of these discrepancies. Better agreement is found with the temperatures derived from UV data, but large differences are found for our two hottest objects: \xcma \ and \bcep. As regards the surface gravities, our values are identical, within the errors, to those presented by Heynderickx \etal (\cite{heynderickx}).

\begin{figure*}
\centering
\includegraphics[width=12.0cm]{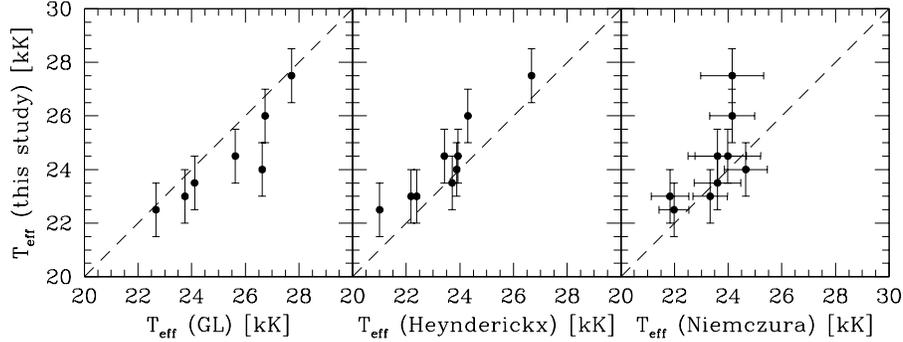}
\caption{Comparison between our $T_{\rm eff}$ values and literature data.}
\label{fig_teff}
\end{figure*}

A variation $\Delta T_{\rm eff}$$\sim$1000 K and $\Delta \log g$$\sim$0.1 dex is inferred during the radial pulsation cycle of \xcma. This is comparable to the uncertainties, but the changes in the EWs of temperature-sensitive lines (Table~\ref{tab_ews}) are such that the temperature variations are clearly real. Furthermore, Beeckmans \& Burger (\cite{beeckmans_burger}) obtained very similar results from UV colour indices: $\Delta T_{\rm eff}$=1400$\pm$430 K. This can be compared to values obtained by various means for other \bc \ stars dominated by a large-amplitude radial mode: e.g., $\Delta T_{\rm eff}$=4000$\pm$2000 K and $\Delta \log g$=0.7$\pm$0.4 dex for $\sigma$ Sco (van der Linden \& Butler \cite{vanderlinden_butler}), $\Delta T_{\rm eff}$=900$\pm$120 K for \voph \ (Morton \& Hansen \cite{morton_hansen}) or $\Delta T_{\rm eff}$$\sim$1200 K for \dcet \ (De Ridder \etal \cite{deridder02}).

\subsection{Stellar abundances}\label{sect_results_abundances} 
The mean NLTE abundances are given in Table~\ref{tab_abundances}, along with the resulting metallicity, $Z$. As a consistency check, we have repeated the abundance analysis of \xcma \ for the two phase ranges considered, finding nearly identical results in both cases. A detailed abundance analysis has been performed for seven stars in our sample by GL. However, a detailed comparison is probably not warranted for the following reasons: (a) the two sets of atmospheric parameters adopted are systematically different, with their temperature scale being likely too hot (see discussion above); (b) the NLTE corrections applied are based on atmospheric models with an inadequate treatment of metal line blanketing (Gold \cite{gold}). Nitrogen being of particular relevance (see below), we simply note the perhaps fortuitous close agreement between the NLTE abundances, with differences never exceeding 0.2 dex. Ignoring previous LTE studies (e.g. Watson \cite{watson71}; Peters \cite{peters}), the same quantitative agreement is found for \gpeg, \dcet \ and \voph \ with other NLTE CNO abundance determinations in the literature (Andrievsky \etal \cite{andrievsky}; Korotin \etal \cite{korotin99a},b,c; Neiner \etal \cite{neiner}).

\begin{table*}
\centering
\caption{Mean NLTE abundances, along with the total 1-$\sigma$ uncertainties, $\sigma_T$ (by convention, $\log \epsilon$[H]=12). The number of used lines is given in brackets. For comparison purposes, the last four columns give: the unweighted mean values for our sample (Mean), typical values found for OB dwarfs in the solar neighbourhood (Daflon \& Cunha \cite{daflon_cunha}; OB stars), the standard solar composition of Grevesse \& Sauval (\cite{grevesse_sauval}; Sun 1-D), and finally solar abundances recently derived from 3-D hydrodynamical models (Asplund \etal \cite{asplund}; Sun 3-D). The metallicity, $Z$, is calculated assuming that the abundances of the elements not under study is solar (Grevesse \& Sauval \cite{grevesse_sauval}). We define [N/C] and [N/O] as $\log$[$\epsilon$(N)/$\epsilon$(C)] and $\log$[$\epsilon$(N)/$\epsilon$(O)], respectively.}
\label{tab_abundances}
\scriptsize
\begin{tabular}{lllllllll} \hline\hline
 & \multicolumn{1}{c}{\gpeg} & \multicolumn{1}{c}{\dcet} & \multicolumn{1}{c}{\neri} & \multicolumn{1}{c}{\bcma} & \multicolumn{2}{c}{\xcma} & \multicolumn{1}{c}{\vcen} & \multicolumn{1}{c}{\voph}\\
 &&& &      & \multicolumn{1}{c}{Max EWs} & \multicolumn{1}{c}{Min EWs} & & \\\hline
He/H                & 0.079$\pm$0.025 (9) & 0.096$\pm$0.024 (10) & 0.076$\pm$0.024 (8) & 0.070$\pm$0.028 (9) & 0.098$\pm$0.017 (10) & 0.099$\pm$0.016 (10) & 0.078$\pm$0.024 (10) & 0.118$\pm$0.032 (9)\\
$\log \epsilon$(C)  & 8.20$\pm$0.05 (8)   & 8.09$\pm$0.08 (6)    & 8.24$\pm$0.12 (10)  & 8.16$\pm$0.11 (9)   & 8.19$\pm$0.11 (9)    & 8.17$\pm$0.12 (9)    & 8.34$\pm$0.13 (8)    & 8.21$\pm$0.07 (4) \\ 
$\log \epsilon$(N)  & 7.58$\pm$0.11 (23)  & 8.05$\pm$0.11 (26)   & 7.87$\pm$0.09 (18)  & 7.59$\pm$0.14 (26)  & 8.00$\pm$0.16 (34)   & 7.99$\pm$0.18 (34)   & 7.73$\pm$0.10 (21)   & 7.99$\pm$0.17 (10)\\ 
$\log \epsilon$(O)  & 8.43$\pm$0.28 (21)  & 8.45$\pm$0.26 (19)   & 8.51$\pm$0.24 (19)  & 8.62$\pm$0.18 (30)  & 8.59$\pm$0.16 (34)   & 8.59$\pm$0.17 (34)   & 8.49$\pm$0.24 (25)   & 8.39$\pm$0.30 (14) \\ 
$\log \epsilon$(Mg) & 7.44$\pm$0.25 (1)   & 7.52$\pm$0.29 (1)    & 7.38$\pm$0.24 (1)   & 7.30$\pm$0.18 (2)   & 7.37$\pm$0.20 (1)    & 7.38$\pm$0.20 (1)    & 7.41$\pm$0.19 (3)    & 7.35$\pm$0.32 (1)\\ 
$\log \epsilon$(Al) & 6.12$\pm$0.16 (4)   & 6.13$\pm$0.21 (4)    & 6.08$\pm$0.16 (4)   & 6.00$\pm$0.15 (4)   & 6.16$\pm$0.21 (4)    & 6.16$\pm$0.23 (4)    & 6.13$\pm$0.11 (4)    & 6.09$\pm$0.24 (2) \\ 
$\log \epsilon$(Si) & 7.19$\pm$0.29 (6)   & 7.28$\pm$0.29 (6)    & 7.21$\pm$0.26 (9)   & 7.17$\pm$0.23 (9)   & 7.14$\pm$0.23 (4)    & 7.12$\pm$0.19 (4)    & 7.14$\pm$0.19 (9)    & 7.16$\pm$0.37 (6)  \\ 
$\log \epsilon$(S)  & 7.22$\pm$0.20 (14)  & 7.26$\pm$0.24 (4)    & 7.32$\pm$0.22 (5)   & 7.14$\pm$0.25 (2)   & 6.98$\pm$0.11 (2)    & 7.00$\pm$0.21 (2)    & 7.39$\pm$0.26 (2)    & 7.32$\pm$0.21 (5)\\
$\log \epsilon$(Fe) & 7.25$\pm$0.16 (28)      & 7.32$\pm$0.18 (21)       & 7.36$\pm$0.19 (17)      & 7.17$\pm$0.19 (25)      & 7.31$\pm$0.19 (32)         &     7.28$\pm$0.25 (32)         &     7.30$\pm$0.09 (16)         & 7.37$\pm$0.21 (9)   \\
$Z$                 & 0.0091$\pm$0.0021         & 0.0100$\pm$0.0021          & 0.0105$\pm$0.0022         & 0.0104$\pm$0.0021         & 0.0110$\pm$0.0018         &     0.0108$\pm$0.0020            &     0.0105$\pm$0.0022            & 0.0097$\pm$0.0022   \\\hline 
${\rm [N/C]}$       & --0.62$\pm$0.12     & --0.04$\pm$0.14      & --0.37$\pm$0.15     & --0.57$\pm$0.18     & --0.19$\pm$0.20      & --0.18$\pm$0.22      & --0.61$\pm$0.17      & --0.22$\pm$0.19\\
${\rm [N/O]}$       & --0.85$\pm$0.30     & --0.40$\pm$0.29      & --0.64$\pm$0.26     & --1.03$\pm$0.23     & --0.59$\pm$0.23      & --0.60$\pm$0.25      & --0.76$\pm$0.26      & --0.40$\pm$0.35\\
\hline
\end{tabular}
\end{table*}

\begin{table*}
\centering
\scriptsize
\begin{tabular}{lllcccc} \hline\hline
 & \multicolumn{1}{c}{\bcep} & \multicolumn{1}{c}{\12lac} & \multicolumn{1}{c}{Mean} & OB stars & Sun 1-D & Sun 3-D\\
 &                          &                              &                             &  \\\hline
He/H                & 0.078$\pm$0.028 (8) & 0.075$\pm$0.031 (8) & 0.085$\pm$0.015   & 0.10$^{a}$    & 0.085  & 0.085\\
$\log \epsilon$(C)  & 8.02$\pm$0.10 (11)  & 8.22$\pm$0.12 (6)   & 8.18$\pm$0.09     & $\sim$8.2     & 8.52   & 8.39\\ 
$\log \epsilon$(N)  & 7.91$\pm$0.13 (19)  & 7.64$\pm$0.18 (17)  & 7.82$\pm$0.19     & $\sim$7.6     & 7.92   & 7.78\\ 
$\log \epsilon$(O)  & 8.47$\pm$0.14 (30)  & 8.42$\pm$0.23 (22)  & 8.49$\pm$0.08     & $\sim$8.5     & 8.83   & 8.66\\ 
$\log \epsilon$(Mg) & 7.31$\pm$0.21  (1)  & 7.29$\pm$0.23 (1)   & 7.37$\pm$0.07     & $\sim$7.4     & 7.58   & 7.53\\ 
$\log \epsilon$(Al) & 6.02$\pm$0.16  (4)  & 6.11$\pm$0.17 (3)   & 6.09$\pm$0.05     & $\sim$6.1     & 6.47   & 6.37\\ 
$\log \epsilon$(Si) & 7.11$\pm$0.23  (8)  & 7.11$\pm$0.27 (6)   & 7.17$\pm$0.05     & $\sim$7.2     & 7.55   & 7.51\\ 
$\log \epsilon$(S)  & 7.14$\pm$0.37  (2)  & 7.10$\pm$0.31 (2)   & 7.21$\pm$0.13     & $\sim$7.2     & 7.33   & 7.14\\ 
$\log \epsilon$(Fe) & 7.24$\pm$0.23 (23)  & 7.30$\pm$0.20 (22)  & 7.29$\pm$0.06     & $\sim$7.4$^b$ & 7.50   & 7.45\\
$Z$                 & 0.0091$\pm$0.0013   & 0.0089$\pm$0.0018   & 0.0099$\pm$0.0007 & $\sim$0.0099  & 0.0172 & 0.0124\\\hline 
${\rm [N/C]}$       & --0.11$\pm$0.17     & --0.58$\pm$0.22     & --0.36$\pm$0.21   & $\sim$--0.6   & --0.60 & --0.61\\
${\rm [N/O]}$       & --0.56$\pm$0.19     & --0.78$\pm$0.29     & --0.67$\pm$0.21   & $\sim$--0.9   & --0.91 & --0.88\\
\hline
\end{tabular}
\begin{flushleft}
$^a$ From Lyubimkov \etal (\cite{lyubimkov04}). The primordial helium abundance is He/H$\sim$0.08 (e.g. Olive \& Skillman \cite{olive_skillman}).\\
$^b$ Mean NLTE abundance of four BA supergiants in the solar vicinity (Przybilla \etal \cite{przybilla06}).
\end{flushleft}
\end{table*}

\section{Discussion} \label{sect_discussion}
\subsection{Neglect of stellar winds} \label{sect_wind}
It is natural to ask to what extent the results presented in this paper are robust against the neglect of the stellar wind, particularly in view of the fact that the inferred microturbulent velocities are in some cases comparable to the sound velocity in the \ion{O}{ii} line-forming regions (recall that an overestimate of this parameter would lead to spuriously low abundances). As expected, relaxing the assumption of LTE leads to a decrease of this quantity, but relatively large values are still obtained with a full NLTE treatment. In the case of \bcma, for instance, $\xi$ decreases from 18 to only 14 km s$^{-1}$ when NLTE is enforced. Furthermore, although our sample is admittedly small and the range in surface gravity limited, there is an indication for higher microturbulent velocities in the more evolved objects (see also Kilian \cite{kilian92} and Daflon \etal \cite{daflon04}). It is tempting to interpret these high microturbulences as an artefact of our assumption of a hydrostatic photosphere (e.g. Lamers \& Achmad \cite{lamers_achmad}). However, there is ample evidence in the literature indicating that high microturbulent velocities are needed, even when state-of-the-art, ''unified'' codes accounting for mass loss and spherical extension are used (e.g. Mokiem \etal \cite{mokiem}).

A number of studies have investigated in recent years the differences between the atmospheric parameters and/or abundances of early-type stars obtained using the plane-parallel code TLUSTY (Hubeny \& Lanz \cite{hubeny_lanz}) and the unified codes CMFGEN/FASTWIND (Hillier \& Miller \cite{hillier_miller}; Puls \etal \cite{puls}). This comparison was performed for a small sample of Galactic/SMC B-type supergiants (Urbaneja \etal \cite{urbaneja}; Trundle \etal \cite{trundle}; Dufton \etal \cite{dufton05}) and Galactic O-type dwarfs (Bouret \etal \cite{bouret05}; Martins \etal \cite{martins}). The differences were claimed to be of the order of the uncertainties and hence of no significance. These consistency checks support the validity of our analysis, especially considering that weaker winds by at least two orders of magnitude are expected in our sample. Perhaps of more relevance is the fact that the same conclusion was reached for some late O-type dwarfs in the SMC having mass-loss rates comparable to what is anticipated for \bc \ stars ($\dot{M}$ down to 10$^{-9}$--10$^{-10}$ M$_{\sun}$ yr$^{-1}$; Bouret \etal \cite{bouret03}). A very weak wind was derived for the most evolved star in our sample, \bcma, from soft X-ray observations: $\dot{M}$=6 $\times$ 10$^{-9}$ M$_{\sun}$ yr$^{-1}$ (Drew \etal \cite{drew}). 

In summary, although forthcoming analyses of our programme stars using unified models would obviously be desirable, the choice of the line formation code used is very unlikely to affect our conclusions about the abundance patterns and does not alleviate our concerns regarding the high microturbulent velocities.
 
\subsection{Abundance patterns} \label{sect_patterns}
As can be seen in Table~\ref{tab_abundances}, the abundances of He, C, O, Mg, Al, Si and Fe span a limited range in the studied stars (less than 0.3 dex). This is comparable to the total uncertainties and indicates that the abundances of these species are remarkably uniform among our sample. The homogeneous nature of our sample in terms of spectral type implies that our results should be fairly insensitive to several sources of systematic errors. Nevertheless, we verified that no significant trends exist between the abundances and $T_{\rm eff}$. This important check supports the reliability of our NLTE corrections and temperature scale and in turn partly explains the small scatter within our sample. Nitrogen and sulphur exhibit a significantly larger spread of the order of 0.4--0.5 dex. The S abundances are derived from very few lines (sometimes of different ionization stages), which suggests that this scatter may not have a physical basis. On the contrary, we will argue below that the star-to-star N variations are real and may be interpreted in terms of mixing processes (Sect.~\ref{sect_mixing}).

One issue we wish to address with the present study is the possible existence of a dichotomy between the chemical properties of the \bc \ stars and those of non-pulsating early B-type stars. In the following we shall use the mean abundance of OB stars in the solar neighbourhood as a baseline for comparison purposes. The final results of a comprehensive, homogeneous abundance study of about 90 late O- to early B-type Galactic stars on (or close to) the main sequence have recently been presented by Daflon \& Cunha (\cite{daflon_cunha}). The analysis is similar in several aspects to our study (use of DETAIL/SURFACE and Kurucz models), a fact which reduces the systematic errors. The typical abundances at the solar galactocentric distance (all our stars lie within 1 kpc) are quoted for all the elements in Table~\ref{tab_abundances}, along with the mean values for our sample.\footnote{It is customary in seismic studies to use the standard solar mixtures of Grevesse \& Sauval (\cite{grevesse_sauval}) or more recent values based on 3-D hydrodynamical models, as summarized by Asplund \etal (\cite{asplund}). These values are also quoted for convenience in Table~\ref{tab_abundances}.} A direct comparison shows a near-perfect agreement for He, C, O, Mg, Al, Si and S. Our mean LTE iron abundance ($\sim$7.18 dex) is in reasonable agreement with the value of $\sim$7.3 dex estimated by Daflon \etal (\cite{daflon99}, \cite{daflon01a}) for early B-type dwarfs. However, we prefer to compare our results to the NLTE Fe abundances of four relatively nearby ($d$ $\la$ 3 kpc) late-B to early-A supergiants obtained by Przybilla \etal (\cite{przybilla06}) using a set of optical \ion{Fe}{ii} lines and the model atom developed by Becker (\cite{becker98}). The Fe abundance is not altered during the evolution of massive stars, allowing for a meaningful comparison. Our mean NLTE abundance is slightly lower than the value quoted by Przybilla \etal (\cite{przybilla06}), but is still compatible when taking the errors into account (Table~\ref{tab_abundances}). 

\subsection{Evidence for deep mixing} \label{sect_mixing}
As stated above, nitrogen is the chemical species exhibiting the largest star-to-star spread. This may indicate the existence in some stars of the products of (incomplete) CNO-cycle burning brought to the surface by deep internal mixing. Evolutionary effects are known to greatly enhance the photospheric N abundance in BA supergiants, but fast rotation leading to meridional currents and shear instabilities in the interior is widely believed to be responsible for mixing prior to the first dredge-up in single early-type stars. The positions of the programme stars in the HR diagram are displayed in Fig.~\ref{fig_hr}. This shows that our targets are largely unevolved and have masses ranging from about 9 to 20 M$_{\sun}$. The most recent evolutionary models including rotation suggest that fast-rotating B stars in this mass range could display a measurable N excess of the order of a few tenths of a dex at the end of the main sequence. This should be accompanied by a slight C depletion at the 0.1 dex level, which is comparable to our uncertainties and therefore barely detectable. On the other hand, He and O should remain virtually unaffected (Heger \& Langer \cite{heger_langer}; Meynet \& Maeder \cite{meynet_maeder03}). The signature of deep mixing in the stellar interior can be best revealed by examining the abundance ratios between N, C and O ([N/C] and [N/O]; see Table~\ref{tab_abundances}). This shows that \gpeg, \bcma, \vcen \ and \12lac  have ratios typical of both nearby OB stars and the Sun, whereas \dcet, \xcma, \voph, \bcep \ and to a much lesser extent \neri \ display higher values by up to 0.6 dex (a factor 4). The two subsamples of N-normal and N-rich stars are separated by about 0.5 dex in terms of the [N/C] ratio (which is the most robust diagnostic for an N enrichment), i.e. about three times the typical uncertainty in the determination of this quantity (Table~\ref{tab_abundances}). The exact choice of the chemical species used to derive the microturbulence (oxygen or nitrogen) has little impact on these conclusions, as the [N/C] and [N/O] ratios are affected by at most 0.12 dex in the case of \xcma \ and \bcep \ (see Table~\ref{tab_micro}).

\begin{figure}
\centering
\includegraphics[width=8.5cm]{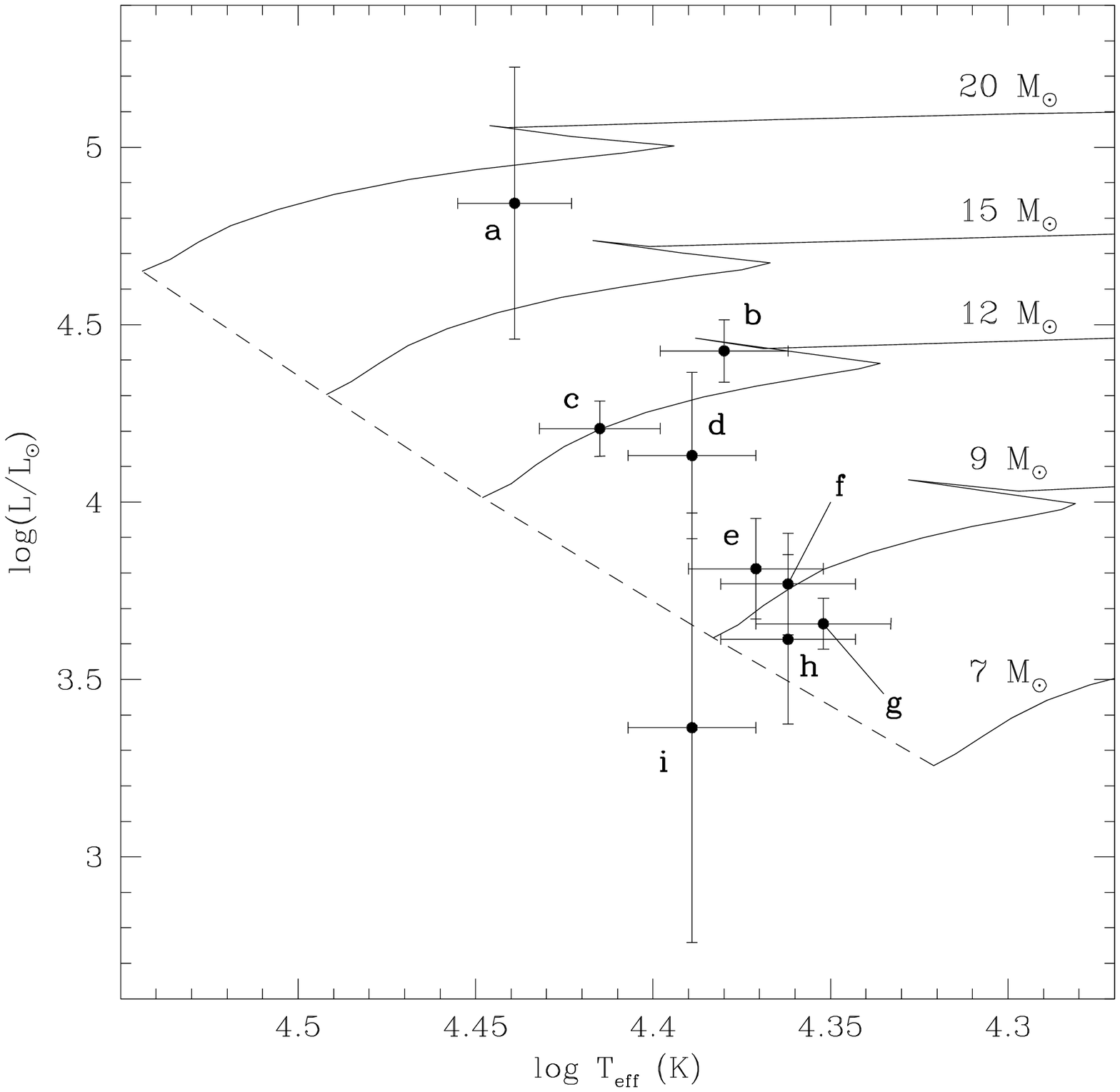}
\caption{Position of the programme stars in the HR diagram: ({\em a}) \xcma, ({\em b}) \bcma, ({\em c}) \bcep, ({\em d}) \12lac, ({\em e}) \neri, ({\em f}) \dcet, ({\em g}) \gpeg, ({\em h}) \voph \ and ({\em i}) \vcen. The luminosities were computed using {\em Hipparcos} parallaxes and the bolometric corrections of Flower (\cite{flower}). The extinction in the $V$ band, $A_V$, was derived from the theoretical ($B-V$) colour indices of Bessell \etal (\cite{bessell}). The evolutionary tracks for solar metallicity and without rotation are taken from Schaller \etal (\cite{schaller}). The initial masses are indicated on the right-hand side of this figure. The ZAMS is shown as a dashed line.} 
\label{fig_hr}
\end{figure}

Such abundance patterns are in line with the predictions of the theoretical models cited above and support the idea that some objects are indeed experiencing substantial rotationally-induced mixing. Boron abundances are available for seven stars in our sample and can be used to further examine the relevance of this interpretation (Proffitt \& Quigley \cite{proffitt_quigley}; Venn \etal \cite{venn02}; Mendel \etal \cite{mendel}). This fragile element is destroyed by proton capture at relatively low temperatures ($T$ $\sim$ 4 $\times$ 10$^6$ K) and is already completely depleted throughout most of the stellar envelope during the very early phases of evolution off the zero-age main sequence (ZAMS). The boron surviving in the outer stellar layers will quickly be transported downwards and be destroyed at the onset of rotational mixing. This will be followed by an increase of the nitrogen content as core-processed gas is being gradually mixed to the surface. As a result, one expects the N-rich stars to be most severely B depleted and, conversely, the stars with little B depletion to have near-solar N abundances. This is consistent with the situation depicted in Fig.~\ref{fig_boron}. 

\begin{figure*}
\centering
\includegraphics[width=17cm]{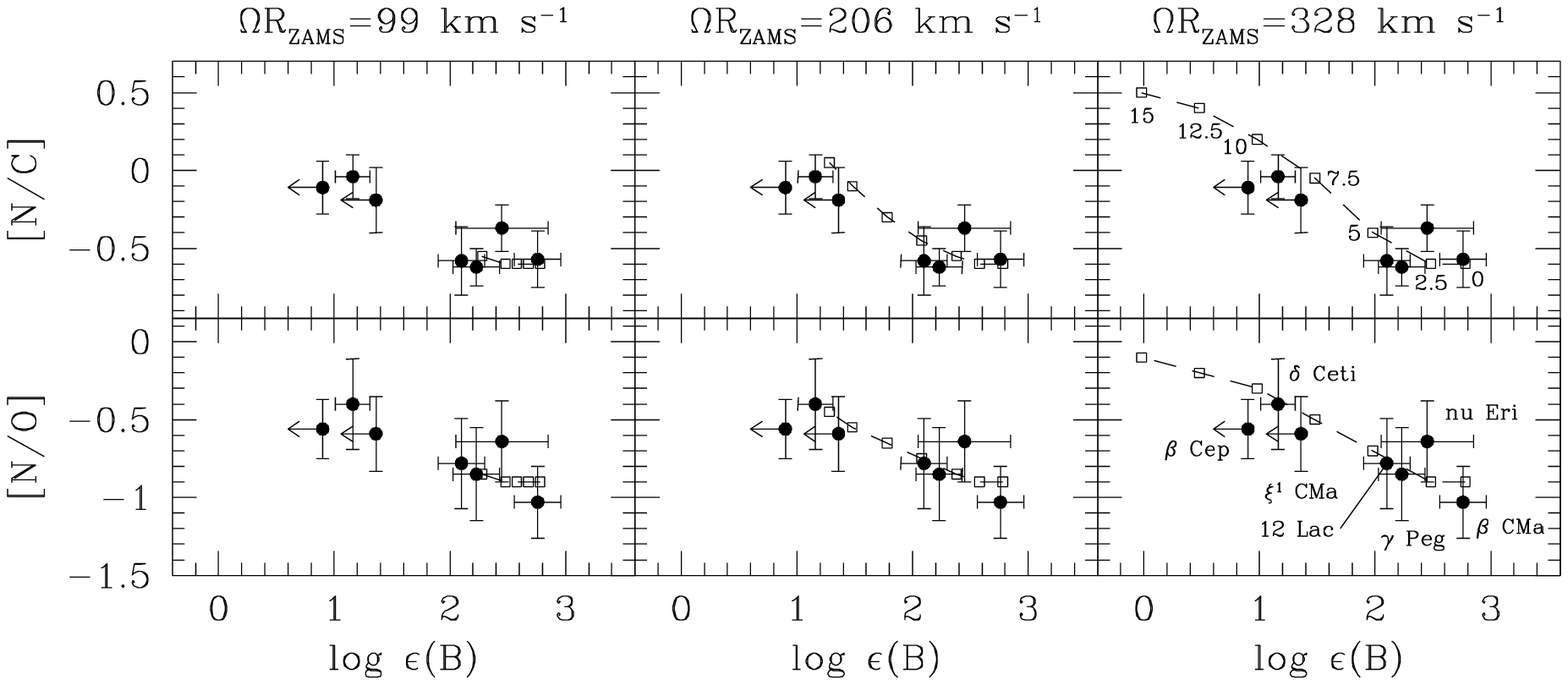}
\caption{[N/C] and [N/O] ratios as a function of the NLTE B abundances. The boron data are primarily taken from Proffitt \& Quigley (\cite{proffitt_quigley}), but have been complemented in the case of \dcet, \bcep \ and \12lac by more stringent estimates from the literature (Venn \etal \cite{venn02}; Mendel \etal \cite{mendel}). The arrows denote upper limits. The abundance data are compared to the theoretical predictions of Heger \& Langer (\cite{heger_langer}) for a 12 M$_{\sun}$ star and three different values of the equatorial rotational velocity on the ZAMS: $\Omega R_{\rm ZAMS}$=99 ({\em left-hand panels}), 206 ({\em middle panels}) and 328 km s$^{-1}$ ({\em right-hand panels}). The locus in each panel ({\em dashed line and open squares}) defines an age sequence with the time elapsed from the ZAMS increasing leftwards: from $t$=0 to 15 Myrs (0 to 12.5 Myrs for $\Omega R_{\rm ZAMS}$=99 km s$^{-1}$) by steps of 2.5 Myrs (see upper-right panel). The initial [N/C], [N/O], and boron abundances at $t$=0 have been taken as: --0.6, --0.9 (Table~\ref{tab_abundances}) and $\log \epsilon$(B)=2.78$\pm$0.04 (meteoritic value; Zhai \& Shaw \cite{zhai_shaw}), respectively.}
\label{fig_boron}
\end{figure*}

However, high initial rotational velocities are required to reproduce the observed abundance patterns. For instance, an initial rotational velocity on the ZAMS, $\Omega R_{\rm ZAMS}$, of about 200 km s$^{-1}$ is necessary to account for the N excess in \dcet \ (with an age of about 14 Myrs according to evolutionary tracks; Fig.~\ref{fig_hr}). The loss of angular momentum along the main sequence leads to a decrease of the rotation rate, but $\Omega R$ is still expected to be in this case of the order of 150 km s$^{-1}$ at the end of core H burning (Heger \& Langer \cite{heger_langer}; Meynet \& Maeder \cite{meynet_maeder03}). This is in excellent agreement with the observations of early-B dwarfs in young open clusters, which reveal only a slow decline of the rotation rate from $\sim$170 km s$^{-1}$ on the ZAMS to $\sim$130 km s$^{-1}$ at the end of core-hydrogen burning (Huang \& Gies \cite{huang_gies}). Most stars enriched in nitrogen have just evolved off the main sequence (Fig.~\ref{fig_hr}), such that reproducing the observed abundance ratios requires high rotational velocities. This is in stark contrast with the fact that none of our targets is apparently rapidly rotating ($\Omega R\sin i$ $<$ 65 km s$^{-1}$; Table~\ref{tab_parameters}). 

Comparisons between the surface abundances of the products of nucleosynthesis and the predictions of models accounting for mass loss and rotation are usually performed in a statistical sense (e.g. Herrero \& Lennon \cite{herrero_lennon}) and rely on the determination of CNO abundances for a large sample of stars only having an estimate of their {\em projected} rotational velocity. One of the key features of this study is that the {\em true} rotation rates, $\Omega R$, have been estimated for several of our targets by a detailed modelling of either the line-profile variations or the pulsation spectrum (see Table~\ref{tab_parameters}), offering an opportunity to constrain the models in a more quantitative way. First, seismic modelling of \neri \ and \vcen \ has revealed a non-rigid rotation profile with very low equatorial velocities of about 6 and 2 km s$^{-1}$, respectively (Dupret \etal \cite{dupret}; Pamyatnykh \etal \cite{pamyatnykh04}). {\em MOST} observations of \dcet \ also lead to a likely rotational velocity of about 28 km s$^{-1}$, or half this value depending on the exact mode identification (Aerts \etal \cite{aerts06}). Second, modelling of the line-profile variations yields $\Omega R$=32 and 45 km s$^{-1}$ in \bcma \ and \12lac, respectively (Desmet \etal \cite{desmet}; Aerts \cite{aerts96}). In addition, assuming that the periodicity of the changes affecting the UV lines can be identified with the rotational period (Neiner \etal \cite{neiner}; Henrichs \etal \cite{henrichs}), leads to $\Omega R$$\sim$56 and 26 km s$^{-1}$ in \voph \ and \bcep, respectively. Although this quantity is unfortunately unknown for \xcma, this shows that at least three N-rich stars (\dcet, \voph \ and \bcep) are intrinsically slow rotators and should not exhibit any abundance peculiarities according to evolutionary models.

Meridional currents are believed to be the most important mechanism governing the loss of angular momentum in (non-magnetic) B stars experiencing little mass loss, while shear mixing arising from differential rotation largely controls the transport of the chemical elements (Meynet \& Maeder \cite{meynet_maeder00}). The observed discrepancy between the observations and the theoretical predictions could  imply that the treatment of these two physical processes is still approximate in evolutionary models including rotation. It is also conceivable, for instance, that the internal rotation law, $\Omega$($r$), is steeper than currently assumed. A decline of the rotational velocity by a factor 3--5 from the core to the surface has been established from asteroseismological studies in two of our targets: \neri \ (Pamyatnykh \etal \cite{pamyatnykh04}) and \vcen \ (Dupret \etal \cite{dupret}). This appears broadly consistent with the predictions of theoretical models, but this piece of information is unfortunately lacking for the N-rich stars. Additional uncertainties arise from our very limited knowledge of the redistribution of angular momentum in massive pulsating stars (see, e.g. Ando \cite{ando}; Lee \& Saio \cite{lee_saio}). From the observational side, there appears to be an excess of slow rotators among the \bc \ class compared to the global B dwarf population, but an interpretation of this result is made difficult by the fact that stars with low $\Omega R\sin i$ values tend to exhibit photometric changes of larger amplitudes and are hence more easily recognizable as \bc \ variables (Stankov \& Handler \cite{stankov_handler}).

It may be regarded as suggestive that 3 out of the 4 stars with an N excess possess a magnetic field with a longitudinal strength of $\sim$300 G (\xcma; Hubrig \etal \cite{hubrig}) and $\sim$100 G (\voph: Neiner \etal \cite{neiner}; \bcep: Henrichs \etal \cite{henrichs}), whereas similar observations for 3 stars with normal abundances (\neri, \bcma \ and \vcen) did not yield any detection (Schnerr \etal \cite{schnerr}; Hubrig \etal \cite{hubrig}). No field has been so far reported in \dcet \ (Rudy \& Kemp \cite{rudy_kemp}), but much more sensitive observations are being planned (Briquet, private communication). In the models of Maeder \& Meynet (\cite{maeder_meynet}), magnetic fields tend to enforce near-solid body rotation and therefore to suppress shear mixing. However, meridional circulation is in this case even more efficient at transporting chemical elements from the interior to the surface. Globally, one expects higher N surface abundances, but also the star to rotate faster. The inclusion of magnetic fields does not seem at first glance able to account for the observational data, but the models are still in their infancy and the theoretical predictions sensitive to some ill-defined quantities (e.g. diffusion coefficients). In this regard, we note that the predicted surface enrichments may strongly differ depending on the model considered (Maeder \& Meynet \cite{maeder_meynet}; Heger \etal \cite{heger}). Furthermore, dynamo action in the radiative layers is assumed to be driven by the so-called Tayler-Spruit instabilities (Spruit \cite{spruit}), the existence of which is not firmly established and, in any case, poorly constrained from an observational point of view. Perhaps more importantly, complex field geometries are expected in the framework of this model, whereas \voph \ and  \bcep \ possess a global dipole field (Neiner \etal \cite{neiner}; Henrichs \etal \cite{henrichs}). This seems to argue for a magnetic field of fossil origin at least in these two objects. While it might be possible to explain in this context the existence of slowly-rotating magnetic B-type dwarfs depending on their pre-main sequence history (St\c{e}pie\'n \cite{stepien}), a nitrogen enrichment at the surface of such objects remains unexpected. Alternatively, one may postulate that these stars have entered the ZAMS with high rotational velocities, but have experienced a dramatic loss of angular momentum along the main sequence because of magnetic braking. However, this interpretation is very unlikely (see Donati \etal \cite{donati} in the case of \bcep).

One issue which needs to be clarified is whether the observed abundance ratios are only intrinsic to the \bc \ stars or are a characteristic feature of early B-type dwarfs in general. We show in Fig.~\ref{fig_CNO} the NLTE [N/C] and [N/O] ratios for our stars and 34 nearby B1--B2 dwarfs, as a function of $\Omega R\sin i$. When gathering the literature data, we restricted ourselves to stars spanning the same $T_{\rm eff}$ and $\log g$ ranges as our programme stars (see Table~\ref{tab_parameters}). In addition, only stars lying at less than about 2 kpc from the Sun were selected to avoid possible [N/C] and [N/O] variations arising from the chemical evolution of the Galaxy (Rolleston \etal \cite{rolleston00}). It turns out that as many as 10 stars in this sample are confirmed \bc \ stars (Stankov \& Handler \cite{stankov_handler}). They are plotted, along with our targets, with different symbols in Fig.~\ref{fig_CNO}. Great care must be exercised when interpreting these data in view of their heterogeneity, although using line ratios minimizes the systematic errors. The magnitude of this problem can be assessed by comparing the [N/C] and [N/O] ratios for the two stars with independent determinations: \object{HD 61068} (GL; Kilian \cite{kilian92}) and \object{HD 216\,916} (GL; Daflon \etal \cite{daflon01a}). The difference is only 0.08 dex on average, but can reach up to 0.19 dex. In spite of the limitations mentioned in Sect.~\ref{sect_results_parameters}, we can also compare our results with GL for the seven stars in common. Once again, the differences are small on average (0.12 dex), but are of the order of 0.3 dex in the most extreme case. Table~\ref{tab_abundances} suggests significantly higher mean [N/C] and [N/O] ratios by about 0.25 dex for our subsample of \bc \ stars compared to nearby, presumably non-pulsating B dwarfs (Daflon \& Cunha \cite{daflon_cunha}). However, this is only marginally the case when considering all the data collected from the various sources in the literature shown in Fig.~\ref{fig_CNO} ($\sim$0.1 dex). This illustrates the fact that comparing the chemical properties of these two populations at these levels is sensitive to the exact definition of the samples (e.g. evolutionary status) and, most importantly, to systematics in the abundance determinations. Confidently establishing a possible dichotomy between the abundance patterns must await the homogeneous analysis of a large sample. Another more subtle difficulty lies in the fact that stars classified as non pulsating might prove to be \bc \ stars when observed under closer scrutiny (see Telting \etal \cite{telting06}). We thus conclude at this stage that there is no convincing evidence for a higher amount of mixing in \bc \ stars compared to the global B dwarf population. 

\begin{figure*}
\centering
\includegraphics[width=12cm]{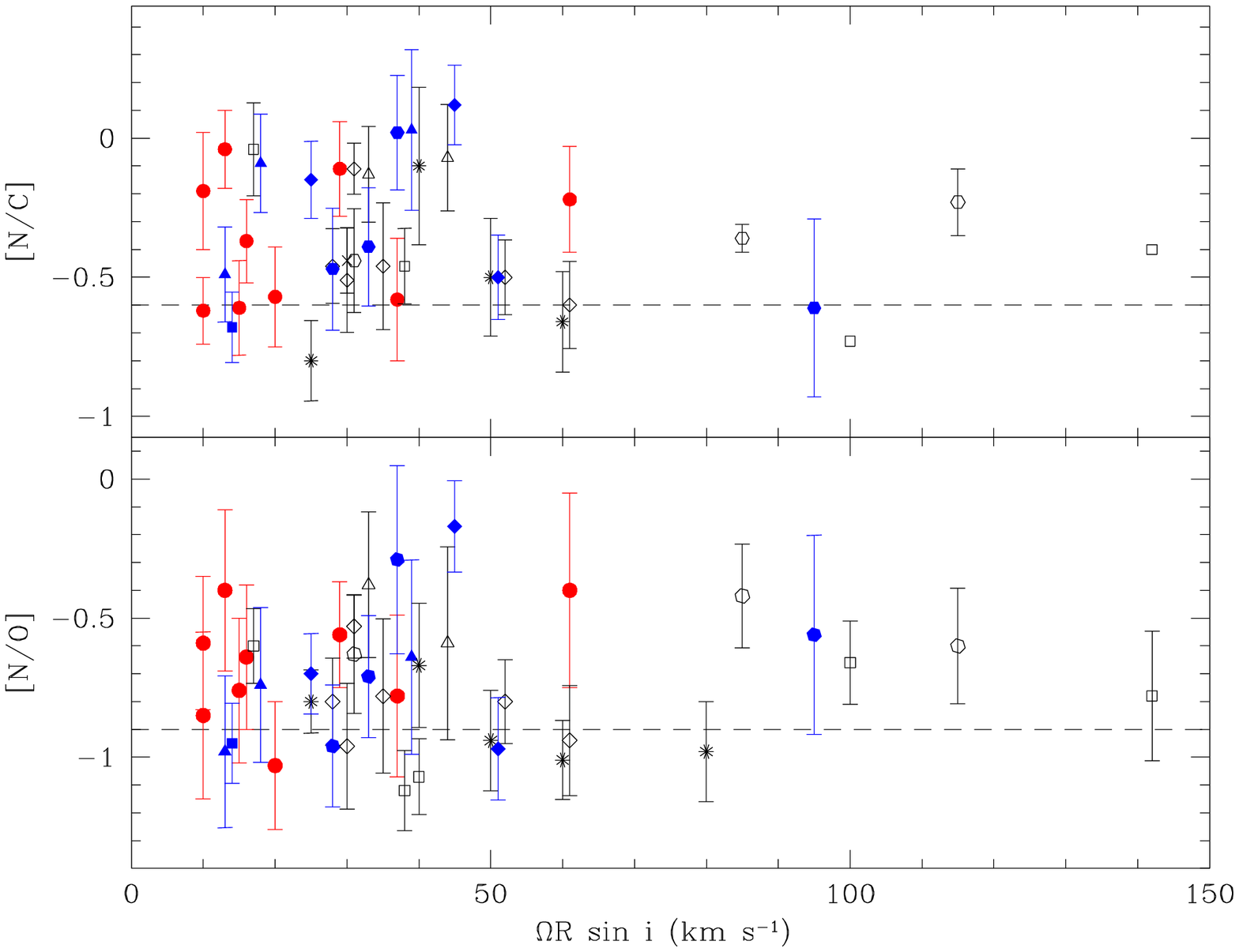}
\caption{[N/C] and [N/O] ratios as a function of $\Omega R\sin i$ for our programme stars ({\em circles}) and a sample of 34 nearby B1--B2 dwarfs: {\em crosses}: Andrievsky \etal (\cite{andrievsky}), {\em squares}: Daflon \etal (\cite{daflon99}, \cite{daflon01a}, b), {\em triangles}: GL, {\em starred}: Gummersbach \etal (\cite{gummersbach}), {\em diamonds}: Kilian (\cite{kilian92}) and Kilian \etal (\cite{kilian94}), {\em hexagons}: Mathys \etal (\cite{mathys}). The filled geometrical symbols denote known \bc\ stars (Stankov \& Handler \cite{stankov_handler}). The dashed lines show the typical ratios in the absence of mixing (Table~\ref{tab_abundances}). The $\Omega R\sin i$ values from Table~\ref{tab_parameters} have been used for our programme stars. The colour coding in the online version of this journal is the following: {\em red}: programme stars, {\em blue}: literature data for other \bc \ stars, {\em black}: presumably non-pulsating B stars.} 
\label{fig_CNO}
\end{figure*}

We note that only the radial pulsator \xcma \ or the stars dominated by a radial mode, \dcet, \voph \ and \bcep, are N-rich. To further examine the potential role played by pulsations, we have investigated a possible link between the N excesses and several pulsational observables (e.g. velocity amplitudes, oscillation frequencies), but found no correlation between these quantities.    

\section{Conclusion and prospects} \label{sect_conclusion}
We have presented a detailed, self-consistent determination of the atmospheric parameters and NLTE abundances of nine prototypical \bc \ stars based on very-high quality spectroscopic data covering the entire oscillation cycle in most cases. This study reveals that the abundance patterns are remarkably similar among our sample, with the exception of nitrogen which is enhanced by a factor 2--4 in four targets. Caution should therefore be exercised when considering apparently slowly-rotating, early B-type dwarfs as good tracers of the present-day nitrogen abundance of the interstellar medium, either in the Galaxy or in the Magellanic Clouds. We interpret this nitrogen enrichment, which is systematically accompanied by a boron depletion, as the result of internal mixing. The existence of stars with normal nitrogen abundances, but with a strong boron depletion (e.g. \12lac; Fig.~\ref{fig_boron}) supports this conclusion (Fliegner \etal \cite{fliegner}). Mass transfer processes in close binaries can mimic such abundance peculiarities (e.g. Vanbeveren \cite{vanbeveren}) and/or to lead to a dramatic spin down because of tidal effects (Huang \& Gies \cite{huang_gies}). However, only two stars in our sample are confirmed members of a (wide) binary system: \gpeg \ (${\cal P}_{\rm orb}$$\sim$370 yr; Chapellier \etal \cite{chapellier}) and \bcep \ (${\cal P}_{\rm orb}$$\sim$90 yr; Pigulski \& Boratyn \cite{pigulski_boratyn}). As \gpeg \ is not N-enriched, this argues against a binary origin. Furthermore, the other stars have all been the subject of long-term, extensive spectroscopic campaigns (except \vcen) and do not present any signs of binarity. For the sake of completeness, we have also computed the peculiar space velocities, $v_{\rm pec}$, of our targets from {\em Hipparcos} data and found that none of the N-enriched stars exceeds the canonical threshold for a runaway status ($v_{\rm pec}$ $\ga$ 40 km s$^{-1}$; Blaauw \cite{blaauw}). Hence there is no evidence for a dynamical kick imparted by the previous explosion of a putative companion as a supernova. Alternatively, the stars could have formed out of interstellar material contaminated by the ejecta of Type II supernovae. This hypothesis can unfortunately not be tested by comparing the derived abundances with those of OB stars in their immediate vicinity, as none of the four stars is known to be part of an open cluster (Stankov \& Handler \cite{stankov_handler}). Likewise, a search for nearby open clusters with abundance data for early-type members in the WEBDA\footnote{The Database for Galactic Open Clusters (WEBDA) is available online at: {\tt http://www.univie.ac.at/webda/.}} database proved inconclusive. Nevertheless, such a scenario is unable to explain the boron data and seems very implausible.

At least three N-rich stars (\dcet, \bcep \ and \voph) are intrinsically slow rotators and should not expose such a large amount of core-processed material at their surfaces according to the most recent evolutionary models accounting for rotation. It should be emphasized that the reliability of the theoretical models has so far mainly been tested from observations of OBA supergiants in the Magellanic Clouds (e.g. Trundle \& Lennon \cite{trundle_lennon}; Heap \etal \cite{heap} for some recent examples). Studies of B dwarfs, such as the present work, are free of several uncertainties which may render a confrontation between theory and observations difficult for such evolved objects: (a) the uncertainties on the mass-loss rates and on the related loss of angular momentum as the star evolves, (b) their ambiguous evolutionary status (pre- or post-red supergiant phase), and (c) the difficulties in estimating the rotation rate of supergiants (e.g. Dufton \etal \cite{dufton06}). One way of reconciling the model predictions with the observations could be to assume that the $\Omega$($r$) law in the interior is much steeper than hitherto assumed. The existence of a magnetic field in most N-rich stars (Sect.~\ref{sect_mixing}) indicates that theoretical models including the effect of magnetic fields are probably more relevant for interpreting our observations. A direct comparison with models assuming that the field is produced via dynamo action in the radiative layers (e.g. Maeder \& Meynet \cite{maeder_meynet}) may not be appropriate, however, as the fields might be of fossil origin in these objects. 

We have determined the abundances of all the chemical species contributing significantly to the global metallicity, $Z$ (except Ne). It is of importance to emphasize that the mild N enrichment observed in some stars has little incidence on the metal content, as this element is only a minor contributor to $Z$ ($\sim$5\%). As can be seen in Table~\ref{tab_abundances}, the metallicity of the studied \bc \ stars is unremarkable and typical of the values observed for early B-type dwarfs in the solar neighbourhood, in accordance with recent works based on UV data (Niemczura \& Daszy\'nska-Daszkiewicz \cite{niemczura_daszynska}). It can be noticed that the Sun is markedly more metal-rich than young, nearby B stars. This long-standing problem still lacks a clear explanation and may constitute a major source of uncertainty when choosing the appropriate metallicity value in oscillation codes. 

Recent multisite, ground-based campaigns have brought a wealth of new high-quality data on \bc \ stars (e.g. Aerts \etal \cite{aerts04a}; Handler \etal \cite{handler}), and much more is expected in the near future thanks to a plethora of space missions (e.g. {\em MOST}, {\em COROT}, {\em Kepler}). The diagnostic power of seismic modelling of early-type stars will hence become enormous, but evolution and oscillation codes unfortunately have difficulties in keeping pace with these dramatic observational developments. In particular, concerns have recently been raised regarding the inability of standard models to reproduce the rich oscillation spectrum of \12lac and \neri. In both cases, the incorporation of gravitational settling and radiative levitation has been claimed as an ad hoc solution to help resolve this problem (Pamyatnykh \etal \cite{pamyatnykh04}; Handler \etal \cite{handler}). We find that both the [N/C] and the [N/O] ratios are about 0.25 dex above solar in \neri \ (Table~\ref{tab_abundances}), a star whose rotation rate only decreases by a factor three between the core and the surface (Pamyatnykh \etal \cite{pamyatnykh04}). This is likely significant and provides some indication for deep mixing. Such an interpretation would challenge the existence of diffusion effects at least in this star, as such microscopic processes are known to be severely hindered, or even completely inhibited, by large-scale fluid motions in the interior. However, P.-O. Bourge and collaborators have called our attention to the fact that radiatively-driven microscopic diffusion, and not rotational mixing, could actually be responsible for the N-excess in \bc \ stars. They have already shown that this process can indeed lead to the accumulation of iron in the transition region of \bc \ stars, thus explaining the excitation of all the observed modes in \12lac and \neri \ (Bourge \& Alecian \cite{bourge_alecian}; Bourge \etal \cite{bourge06a}). Their most recent results show that radiative forces on nitrogen are always greater than those on carbon and oxygen near the surface, thus leading to separation and variable excess of N depending on the stellar parameters (Bourge \etal \cite{bourge06b}). This intriguing possibility opens a new avenue for the interpretation of the abundance peculiarities discussed in this paper.

\begin{acknowledgements}
T. M. acknowledges financial support from the European Space Agency through a Postdoctoral Research Fellow grant and from the Research Council of Leuven University through grant GOA/2003/04. M. B. is Postdoctoral Fellow of the Fund for Scientific Research, Flanders. This work benefited from discussions within the Belgian Asteroseismology Group (BAG). We particularly wish to thank P.-O. Bourge, A. Noels, S. Th\'eado and A. Thoul for sharing with us their results regarding diffusion. The useful comments from an anonymous referee are greatly acknowledged. We are indebted to D.~J.~Lennon and C.~Trundle for allowing us to implement their \ion{Si}{ii} model ion in DETAIL. We are also grateful to the many people involved in the acquisition and reduction of the data presented in this paper, as well as F. Favata for support. The archival ELODIE data have been processed within the PLEINPOT environment. We wish to thank P. Prugniel for his help with PLEINPOT and the ELODIE archives. This research made use of NASA's Astrophysics Data System Bibliographic Services, the SIMBAD database operated at CDS, Strasbourg (France) and the WEBDA database operated at the Institute for Astronomy of the University of Vienna. 
\end{acknowledgements}

\Online

\appendix
\section{EW measurements} \label{sect_ews}
Table~\ref{tab_ews} lists the measured EWs for the studied stars. Our EWs are systematically larger than the values quoted by GL for the seven stars in common, but a similar trend is evident when comparing their values with previous measurements in the literature (see their Fig.~4). This is likely to result from their use of Gaussian fitting instead of the direct integration used in our case. The same conclusion holds for the EWs of \xcma \ presented by Hambly \etal (\cite{hambly}).

\begin{table*}
\centering
\caption{EW measurements in m\AA. A blank may have various causes: the EW was not reliably measurable (spectral features lying in the wings of the Balmer and diffuse \ion{He}{i} lines were avoided); the line fell into echelle inter-order gaps; was considered blended for the relevant temperature range; or yielded a discrepant abundance. The latter point notably applies to the \ion{C}{ii} $\lambda$4267 line which is notoriously known to yield spuriously low abundances (e.g. Lennon \etal \cite{lennon}, although Nieva \& Przybilla \cite{nieva_przybilla} appear to have recently solved this problem).}
\label{tab_ews}
\scriptsize
\begin{tabular}{lrrrrrrrrrr} \hline\hline
Transition$^a$ & \multicolumn{1}{c}{\gpeg} & \multicolumn{1}{c}{\dcet} & \multicolumn{1}{c}{\neri} & \multicolumn{1}{c}{\bcma} & \multicolumn{2}{c}{\xcma} & \multicolumn{1}{c}{\vcen} & \multicolumn{1}{c}{\voph} & \multicolumn{1}{c}{\bcep} & \multicolumn{1}{c}{\12lac}\\
             &  &&&       & \multicolumn{1}{c}{Max EWs} & \multicolumn{1}{c}{Min EWs} & &  & & \\\hline
{\bf \ion{He}{i}} && &  & & & &  &  \\
4026.20 & 1245  & 1366  & 1236  & 1049  & 1010  & 1035  &  1207 & 1546  &  1018 &  1109\\
4387.93 & 838.7 & 846.5 & 778.9 & 648.9 & 676.8 & 678.9 & 741.6 & 1046  & 691.7 & 712.4\\ 
4437.55 & 125.5 & 134.6 &       & 118.7 & 104.9 & 104.7 & 122.7 & 154.1 &  86.1 & 108.1\\
4471.48 & 1238  & 1360  &       &       & 1049  & 1094  &  1256 &       &  1124 &      \\
4713.15 & 261.4 & 274.3 & 290.7 & 293.1 & 261.0 & 258.9 & 276.0 & 291.3 & 242.7 &      \\
4921.93 & 826.9 & 850.2 & 820.0 & 725.1 & 689.0 & 705.0 & 804.3 & 1131  & 724.7 & 764.5\\ 
5015.68 & 289.3 & 290.5 & 336.6 & 352.6 & 298.7 & 297.4 & 296.9 & 388.9 & 274.6 & 330.8\\
5047.74 & 168.3 & 188.9 & 193.3 & 182.8 & 170.5 & 164.0 & 181.7 & 201.3 & 144.8 & 169.2\\
5875.62 &       & 775.6 & 834.1 & 854.4 & 749.6 & 760.0 & 751.3 & 839.0 &       & 777.3\\
6678.15 & 615.7 & 663.5 & 735.3 & 777.9 & 671.3 & 679.1 & 664.5 & 800.1 &       & 709.3\\
{\bf \ion{C}{ii}} &&  & &  & & &  &  & & \\
5133.11 &  64.2 &  52.4 &  69.0 &  57.0 &  42.0 &  36.5 &  70.1 &       &  33.8 &      \\
5137.26 &       &       &  10.2 &   6.5 &   5.1 &   4.1 &       &       &   4.6 &      \\
5139.17 &  12.3 &  10.4 &  12.2 &   9.5 &   7.6 &   5.3 &  18.2 &       &   6.0 &  13.0\\
5143.50 &       &       &       &  27.1 &  20.9 &  17.9 &  35.8 &       &  17.8 &  25.7\\
5145.16 &  54.5 &  49.2 &  65.4 &  68.0 &  40.0 &  37.1 &  59.2 &       &  34.8 &  67.1\\
5151.08 &  34.5 &  29.0 &  43.7 &       &       &       &       &  37.1 &       &      \\
6780.26 &  44.7 &  31.4 &  44.0 &       &       &       &  54.8 &  44.4 &  18.9 &  39.8\\
6783.91 &  47.7 &  40.1 &  51.0 &       &  31.0 &  26.9 &  60.7 &  46.0 &  26.0 &  46.0\\
6787.21 &       &       &       &  11.5 &   7.2 &   6.8 &  19.8 &       &   8.3 &      \\
6791.47 &  18.5 &       &  19.5 &  13.7 &   9.5 &   8.2 &  25.2 &  18.2 &   6.3 &      \\
6798.10 &       &       &   3.7 &       &       &       &       &       &       &      \\
6800.69 &  16.2 &       &  15.9 &  11.0 &   8.0 &   5.6 &       &       &   7.3 &      \\
{\bf \ion{C}{iii}} & & & & &  & &  &  & & \\ 
4186.90 &       &       &       &  15.3 &       &       &       &       &  16.4 &  18.4\\
{\bf \ion{N}{ii}}  & & & & &  &  &  &  & &  \\
3955.85 &       &       &       &       &  45.7 &  40.6 &       &       &       &      \\
3995.00 &       &       &       &       & 124.2 & 115.6 &       &       &       &      \\
4179.67 &   8.9 &  19.7 &       &       &  21.5 &  23.2 &  12.6 &  18.8 &       &      \\
4227.74 &  15.8 &  36.8 &  36.4 &  22.6 &  33.7 &  28.9 &  25.8 &  29.0 &  32.6 &  21.9\\
4236.99 &       &       &       &  43.7 &  65.0 &  60.9 &       &       &       &      \\
4564.76 &       &   3.4 &       &       &       &       &       &       &       &      \\
4607.16 &  34.4 &  66.4 &  61.0 &  42.7 &  59.8 &  50.1 &  51.6 &  57.2 &  57.0 &  50.9\\
4643.09 &  40.3 &  74.1 &  82.2 &  41.6 &  65.5 &  55.4 &  57.7 &       &  63.5 &  43.7\\
4654.53 &   6.8 &  17.1 &  11.3 &       &       &       &       &  12.7 &       &      \\
4667.21 &   4.2 &  11.7 &       &   4.9 &       &       &       &       &   9.6 &      \\
4674.91 &   4.5 &  12.4 &       &       &       &       &       &       &   6.9 &      \\
4774.24 &   2.7 &       &       &       &       &       &       &       &       &      \\
4779.72 &       &       &       &  11.4 &  19.9 &  16.3 &  15.4 &       &  19.8 &      \\
4781.19 &   2.1 &   6.6 &   5.0 &       &       &       &       &       &   4.1 &      \\
4788.14 &  14.8 &  32.7 &  29.3 &  20.0 &  28.2 &  24.1 &  21.2 &       &  28.8 &      \\
4987.38 &   8.6 &  17.9 &  19.4 &  13.0 &  14.5 &  12.6 &   9.4 &  18.0 &  14.5 &  14.0\\
4994.36 &  19.8 &       &  44.2 &  31.6 &  36.0 &  31.5 &  35.1 &  35.1 &  34.8 &      \\
5001.30 &       &       &       & 131.0 & 143.1 & 127.1 & 127.7 &       &       & 128.5\\
5002.70 &  15.0 &  32.5 &  36.2 &  20.6 &  30.1 &  24.3 &  21.7 &       &       &  17.2\\
5005.15 &  51.3 &  81.1 &       &  91.6 &  90.6 &  81.8 &  74.8 &  70.9 &       &  74.6\\
5007.33 &       &       &       &  40.1 &  46.0 &  40.2 &  36.0 &       &  45.7 &  36.2\\
\hline
\end{tabular}
\end{table*}

\addtocounter{table}{-1}
\begin{table*}
\centering
\scriptsize
\begin{tabular}{lrrrrrrrrrr} \hline\hline
Transition$^a$ & \multicolumn{1}{c}{\gpeg} & \multicolumn{1}{c}{\dcet} & \multicolumn{1}{c}{\neri} & \multicolumn{1}{c}{\bcma} & \multicolumn{2}{c}{\xcma} & \multicolumn{1}{c}{\vcen} & \multicolumn{1}{c}{\voph} & \multicolumn{1}{c}{\bcep} & \multicolumn{1}{c}{\12lac}\\
               &&& &      & \multicolumn{1}{c}{Max EWs} & \multicolumn{1}{c}{Min EWs} & & & &\\\hline
{\bf \ion{N}{ii}} (Continued) && & & & &  & & & &  \\
5010.62 &  26.7 &  53.8 &  66.5 &  54.0 &  54.2 &  43.7 &  41.9 &       &       &      \\
5011.31 &       &       &       &       &   7.0 &   4.7 &       &       &       &      \\
5012.04 &       &   8.8 &       &       &  12.5 &  10.8 &       &       &       &      \\
5025.66 &       &  22.0 &       &  19.6 &  17.4 &  15.1 &  14.9 &       &       &  22.5\\
5045.10 &       &       &       &  65.1 &  74.8 &  62.6 &  63.0 &       &       &  65.0\\
5462.58 &       &       &       &       &   8.8 &   8.0 &       &       &       &      \\
5480.05 &   3.9 &  10.3 &  10.8 &       &  11.9 &   9.5 &       &       &       &      \\
5495.65 &  12.9 &  24.0 &  28.3 &  24.8 &  30.2 &  25.6 &  23.1 &       &  30.9 &  26.0\\
5551.92 &       &       &       &       &   4.8 &   4.8 &       &       &       &      \\
5666.63 &       &  84.9 &       &  74.1 &  83.6 &  69.4 &  72.9 &  66.4 &  83.7 &  70.1\\
5676.02 &  32.0 &  63.2 &  71.7 &  58.8 &  72.1 &  61.3 &  58.3 &  53.4 &  69.4 &  54.6\\
5679.56 &  69.1 & 113.7 &       & 130.1 & 125.6 & 109.2 & 108.2 & 100.4 &       & 112.9\\
5686.21 &  24.9 &  52.1 &  55.4 &  38.0 &  45.6 &  36.0 &  42.7 &       &  53.0 &  44.6\\
5710.77 &  25.1 &  51.3 &  50.1 &  40.5 &  49.7 &  41.6 &  39.9 &       &  48.1 &  37.3\\
5747.30 &   8.0 &  22.3 &  15.8 &   8.8 &  17.2 &  14.3 &       &       &       &      \\
5767.45 &       &  12.0 &   7.3 &       &   7.9 &   6.8 &       &       &       &      \\
6242.41 &       &       &       &   5.8 &  12.3 &   9.9 &       &       &  12.7 &      \\
6379.62 &  10.0 &  22.3 &  19.9 &  11.7 &  16.7 &  11.9 &       &       &  18.7 &  22.9\\
{\bf \ion{N}{iii}} & & & & & &  &  & & &  \\
4634.13 &       &       &       &   9.3 &  26.9 &  34.9 &       &       &  18.1 &      \\
{\bf \ion{O}{ii}}  & & & & & &  &  &  & & \\
3912.04 &  45.2 &  54.5 &  79.8 & 129.7 & 117.0 & 109.1 &  83.0 &  50.7 & 100.0 &  99.1\\
4069.75 &  92.8 & 110.4 & 181.3 & 272.9 &       &       & 167.9 &  92.5 & 200.5 & 178.5\\
4072.15 &       &       &       & 203.9 & 144.3 & 139.8 & 103.9 &       &       &      \\
4078.86 &       &       &       &  87.4 &  68.5 &  65.0 &       &       &  52.9 &      \\
4084.88 &       &       &       &       & 107.4 & 103.2 &       &       &       &      \\
4129.31 &       &       &       &  31.1 &  25.6 &  24.4 &       &       &       &      \\
4132.79 &       &       &       & 102.9 &  84.5 &  84.6 &  56.6 &       &  68.7 &      \\
4185.44 &  29.5 &  33.1 &  54.5 &       &  79.7 &  79.2 &  54.4 &  31.2 &  64.3 &  63.4\\
4283.33 &       &       &       &       &       &       &  65.4 &       &       &      \\
4319.78 &       &       &       & 182.8 & 142.1 & 134.6 &       &       & 113.4 &      \\
4366.70 &  58.4 &  61.6 & 118.7 & 175.8 & 136.8 & 130.0 &  91.9 &  56.8 & 118.5 &      \\
4414.88 &  73.9 &  81.5 & 151.7 & 231.5 & 179.6 & 173.2 & 120.0 &  67.6 & 161.4 & 150.0\\
4416.97 &       &       &       & 185.9 & 152.1 & 147.4 & 106.5 &       & 138.4 & 120.3\\
4452.38 &  28.5 &  31.2 &  54.1 &  84.0 &  72.9 &  70.1 &  53.2 &       &  63.6 &  59.8\\
4591.01 &  47.7 &  53.5 &  99.0 & 166.3 & 135.4 & 131.0 &  86.9 &  48.5 & 116.2 & 103.0\\
4596.07 &  48.3 &  51.5 &  91.1 & 152.5 & 133.5 & 127.9 &  86.9 &  48.9 & 112.9 & 104.4\\
4609.79 &       &       &       &       & 108.6 & 110.7 &  77.7 &       &  91.7 &  76.1\\
4641.82 &  72.6 &  85.0 & 163.8 & 249.0 & 186.6 & 180.7 & 121.8 &       & 162.9 & 193.1\\
4649.14 &  90.7 & 101.3 &       &       & 219.1 & 211.2 & 156.0 &  73.4 &       & 223.3\\
4661.64 &  53.5 &  59.3 & 113.0 & 182.1 & 141.5 & 135.1 &  95.7 &  53.2 & 121.0 & 117.5\\
4673.75 &       &       &       &  68.6 &  56.6 &  55.5 &  33.0 &       &  46.0 &  42.7\\
4676.24 &  45.4 &  49.0 &       & 166.3 & 119.6 & 114.7 &  78.2 &  52.0 & 106.6 & 114.3\\
4691.15 &   8.7 &       &  15.1 &  25.1 &  26.7 &  25.4 &       &  10.3 &  20.9 &  19.9\\
4696.36 &  13.0 &  16.1 &  23.5 &  38.6 &  35.8 &  32.2 &       &       &  30.2 &  23.0\\
4701.44 &  13.4 &  14.9 &  24.6 &  41.0 &  43.4 &  41.9 &  28.7 &       &  33.7 &  28.2\\
4705.32 &  42.5 &  46.2 &  80.2 & 136.3 & 114.4 & 111.6 &  80.5 &  33.3 &  97.2 &      \\
4741.70 &   5.1 &   4.7 &       &  15.0 &  16.3 &  15.1 &       &       &  12.4 &      \\
4890.85 &       &       &       &  43.6 &  39.0 &  36.4 &       &       &  33.1 &      \\
4906.82 &  20.4 &  23.2 &  43.1 &  73.4 &  63.1 &  60.4 &  41.9 &  21.4 &  56.7 &  42.7\\
4941.10 &  17.6 &  23.8 &  39.7 &  64.0 &  55.9 &  55.6 &  38.3 &  20.9 &  47.1 &  55.9\\
4943.00 &       &       &       &       &  80.0 &  79.9 &  55.1 &       &  70.1 &  71.1\\
4955.74 &   8.2 &  10.6 &  19.2 &  32.6 &  27.7 &  26.8 &       &       &  23.4 &  19.6\\
5175.99 &       &       &       &  18.9 &       &       &       &       &       &      \\
\hline
\end{tabular}
\end{table*}

\addtocounter{table}{-1}
\begin{table*}
\centering
\scriptsize
\begin{tabular}{lrrrrrrrrrr} \hline\hline
Transition$^a$ & \multicolumn{1}{c}{\gpeg} & \multicolumn{1}{c}{\dcet} & \multicolumn{1}{c}{\neri} & \multicolumn{1}{c}{\bcma} & \multicolumn{2}{c}{\xcma} & \multicolumn{1}{c}{\vcen} & \multicolumn{1}{c}{\voph} & \multicolumn{1}{c}{\bcep} & \multicolumn{1}{c}{\12lac}\\
               &&& &      & \multicolumn{1}{c}{Max EWs} & \multicolumn{1}{c}{Min EWs} & & & &\\\hline
{\bf \ion{O}{ii}} (Continued) && & & & &  & & & &  \\
5190.56 &   7.2 &       &       &       &  28.5 &  26.7 &       &       &       &      \\
5206.71 &       &       &       &  57.8 &  51.8 &  49.7 &  30.9 &       &  42.8 &      \\
6640.99 &       &       &  33.7 &  62.1 &  61.3 &  57.2 &       &       &  47.7 &  36.8\\
6721.36 &       &       &  49.2 &  95.9 &  84.7 &  78.0 &  47.1 &       &  71.0 &      \\
{\bf \ion{Mg}{ii}} & & & & & &    &&  & & \\
4481.20 & 156.1 & 160.5 & 205.8 & 170.3 & 133.7 & 127.8 & 161.8 & 149.0$^b$ & 134.4 & 166.0\\
4739.65 &       &       &       &   3.8 &       &       &       &           &       &      \\
7877.05 &       &       &       &       &       &       &  44.5 &           &       &      \\
7896.20 &       &       &       &       &       &       &  53.8 &           &       &      \\
{\bf \ion{Al}{iii}} &&  &       & & &    &&  & & \\
4149.94 &  34.4 &  33.5 &  46.2 &  28.1 &  26.9 &  23.9 &  37.0 &  27.1 &  19.8 &      \\
4479.93 &  40.3 &  46.6 &  51.7 &  53.5 &  40.7 &  36.7 &  61.4 &       &  41.8 &  54.9\\
4512.56 &  38.2 &  40.6 &  52.7 &  47.7 &  37.2 &  33.0 &  48.2 &  38.1 &  31.7 &  46.0\\
4529.07 &  62.9 &  66.9 &  90.8 &  85.0 &  60.3 &  54.7 &  76.6 &       &  53.8 &  87.6\\
{\bf \ion{Si}{ii}} &&  & & & &    &&  & & \\
4128.05 &       &  34.0 &  61.5 &  33.4 &       &       &  33.1 &  41.1 &      &\\
4130.89 &  35.8 &  40.8 &  59.8 &  25.0 &       &       &  31.1 &       & 10.8 &\\
5056.15 &       &       &  25.3 &       &       &       &       &       &      &\\
6371.37 &  40.1 &  29.8 &  23.4 &       &       &       &  23.2 &       &      &\\
{\bf \ion{Si}{iii}} & & &       & & &    &&  & & \\
4567.84 & 100.7 & 112.9 & 215.5 & 277.1 &       &       & 145.5 & 100.5 & 173.8 & 202.1\\
4574.76 &  67.8 &  77.1 & 139.2 & 180.9 &       &       & 101.9 &  66.5 & 117.7 & 134.7\\
4813.33 &  21.6 &       &  42.5 &  53.8 &  43.4 &  37.4 &  35.5 &  23.1 &  38.1 &  41.1\\
4819.77 &       &       &       &  82.7 &       &       &  62.6 &       &  59.8 &  67.3\\
4829.07 &       &       &  55.9 &  77.1 &  63.7 &  58.2 &  53.9 &  38.3 &  55.4 &      \\
5716.29 &       &       &       &       &       &       &       &       &   4.6 &      \\ 
5739.73 &  72.7 &  82.0 & 174.5 & 223.7 &       &       & 119.5 &  84.7 & 137.3 & 166.6\\
{\bf \ion{Si}{iv}} & & & & & &  &  &  & & \\
4212.40 &       &       &       &   9.5 &  18.8 &  23.0 &       &       &  10.4 &   4.6\\
6701.21 &       &       &       &       &   8.6 &  12.4 &       &       &       &      \\
{\bf \ion{S}{ii}}  &&  & & & &    &&  & & \\
4162.48 &  29.0 &  26.0 &  30.4 &       &       &       &       &  39.5 &       &      \\
4174.13 &  22.5 &  23.1 &       &       &       &       &       &  30.6 &       &      \\
4217.18 &   4.8 &       &   5.4 &       &       &       &       &   5.7 &       &      \\
4230.94 &   4.9 &       &       &       &       &       &       &   5.9 &       &      \\
4269.72 &   4.8 &       &       &       &       &       &       &       &       &      \\
4456.38 &   2.7 &       &       &       &       &       &       &       &       &      \\
4483.43 &   6.7 &       &       &       &       &       &       &       &       &      \\
4486.63 &   2.8 &       &       &       &       &       &       &       &       &      \\
4524.81 &  11.9 &  14.7 &  11.8 &       &       &       &  12.3 &       &       &      \\
4656.76 &  5.3  &       &       &       &       &       &       &       &       &      \\
4668.50 &  4.1  &       &       &       &       &       &       &       &       &      \\
4755.02 &  4.2  &       &       &       &       &       &       &       &       &      \\
4763.31 &  2.3  &       &       &       &       &       &       &       &       &      \\ 
4792.01 &  5.3  &       &   4.7 &       &       &       &       &       &       &      \\
4815.55 &       &  24.6 &  22.6 &  15.3 &       &       &  18.4 &  26.6 &   5.7 &  11.7\\ 
{\bf \ion{S}{iii}} &&  & & & &  &  &  & & \\
4361.53 &       &       &       &  57.0 &  49.4 &  44.7 &       &       &  37.3 &  35.3\\
4527.91 &       &       &       &       &   4.9 &   5.3 &       &       &       &      \\ 
{\bf \ion{Fe}{iii}} &&  & & & &  &  &  & & \\
4005.04 &  17.1 &       &       &       &       &       &       &       &       &  16.9\\
4081.01 &       &  17.4 &       &       &  20.9 &  19.2 &       &       &       & \\
4122.78 &       &  17.2 &       &       &  19.1 &  16.0 &       &       &       & \\
4137.76 &  19.0 &  20.9 &       &       &       &       &  31.2 &       &       & \\ 
\hline
\end{tabular}
\end{table*}

\addtocounter{table}{-1}
\begin{table*}
\centering
\scriptsize
\begin{tabular}{lrrrrrrrrrr} \hline\hline
Transition$^a$ & \multicolumn{1}{c}{\gpeg} & \multicolumn{1}{c}{\dcet} & \multicolumn{1}{c}{\neri} & \multicolumn{1}{c}{\bcma} & \multicolumn{2}{c}{\xcma} & \multicolumn{1}{c}{\vcen} & \multicolumn{1}{c}{\voph} & \multicolumn{1}{c}{\bcep} & \multicolumn{1}{c}{\12lac}\\
               &&& &      & \multicolumn{1}{c}{Max EWs} & \multicolumn{1}{c}{Min EWs} & & & &\\\hline
{\bf \ion{Fe}{iii}} (Continued) && & & & &  & & & &  \\
4139.35 &  17.6 &       &       &       &       &       &  31.4 &       &       & \\ 
4154.96 &   5.3 &   6.0 &       &       &       &       &       &       &       & \\
4164.82 &       &       &       &  66.2 &  56.1 &  51.1 &  58.4 &       &  51.7 &  69.0\\
4166.84 &  13.6 &       &       &       &  16.1 &  14.3 &  20.7 &       &  18.5 &  22.2\\
4222.27 &       &  21.5 &  31.1 &       &       &       &       &       &       & \\ 
4248.77 &   4.8 &       &   7.7 &       &       &       &       &       &       & \\ 
4261.39 &   5.3 &   5.1 &       &       &   4.5 &   4.3 &       &   3.6 &   4.2 &   5.7 \\
4296.85 &  11.3 &       &       &  23.6 &  19.9 &  16.8 &       &  12.1 &  20.0 &  21.9 \\
4310.35 &  15.2 &       &  36.7 &  34.7 &  27.8 &  23.3 &  27.6 &  14.0 &  21.4 &  27.4 \\
4419.60 &  32.7 &  34.2 &  52.4 &  44.0 &  30.6 &  22.5 &  44.8 &  27.8 &       &  38.9 \\
4431.02 &  23.7 &  24.3 &       &       &       &       &       &       &       & \\  
5063.42 &  10.1 &   9.2 &  12.7 &   9.0 &   6.3 &   3.2 &  10.6 &  12.6 &       & \\
5086.70 &       &       &  25.2 &  17.3 &  12.7 &   9.1 &  20.9 &  25.2 &  19.3 &  27.8\\
5127.51 &       &       &       &  54.3 &  44.8 &  33.1 &  61.8 &       &  38.3 & \\
5136.11 &       &       &       &   3.3 &   3.6 &   2.3 &       &       &       & \\
5156.11 &  38.4 &  37.5 &  59.9 &  51.8 &  38.2 &  30.4 &       &  37.6 &  36.4 &  53.2\\
5193.91 &       &  19.0 &       &  24.0 &       &       &       &  16.5 &  16.1 &  29.1\\ 
5218.10 &       &       &       &       &   8.2 &   5.5 &       &       &       & \\
5235.66 &  11.6 &  12.7 &  33.4 &  25.9 &  18.1 &  11.3 &  22.2 &       &  19.3 &  32.6\\
5243.31 &       &       &       &  51.7 &  41.9 &  34.1 &       &       &       &  57.6\\
5276.48 &       &       &       &  32.3 &  25.0 &  19.0 &  26.8 &       &  22.1 &  36.9\\
5282.30 &  16.5 &  20.5 &  40.3 &  38.2 &  27.2 &  19.5 &  34.0 &       &  28.5 &  39.1\\
5284.83 &   5.6 &       &  12.5 &  13.8 &   9.8 &   7.8 &       &       &       & \\
5298.11 &   4.7 &   6.3 &  12.1 &  12.2 &   8.0 &   5.3 &       &       &   8.9 &  12.4\\
5299.93 &  11.5 &  13.3 &  24.0 &  21.8 &  15.6 &  11.0 &  22.3 &       &  15.1 &  28.6\\
5302.60 &  13.2 &  15.6 &       &  25.6 &  19.4 &  15.1 &  20.8 &       &  19.1 &  25.9\\
5306.76 &       &       &       &  18.6 &  13.3 &   9.0 &  14.3 &       &  14.4 &  20.1\\
5310.88 &       &       &       &       &   6.7 &   4.4 &       &       &   7.8 & \\
5322.74 &   3.3 &       &       &       &   5.8 &   5.0 &       &       &   7.6 &   9.1\\
5363.56 &       &       &       &  12.3 &   9.9 &   7.7 &       &       &   8.1 & \\
5460.80 &   7.6 &   8.5 &       &   8.6 &   7.8 &   5.6 &       &       &       & \\
5485.52 &   7.6 &   7.8 &  12.3 &  10.9 &       &       &       &       &   8.1 & \\ 
5573.42 &   8.4 &       &  13.7 &  12.7 &  10.1 &   7.1 &  14.3 &  10.5 &  11.5 &  17.7\\
5744.21 &       &       &       &       &   7.4 &   5.2 &       &       &       & \\
5756.39 &       &       &  11.4 &       &       &       &       &       &       & \\ 
5833.94 &       &       &  79.2 &  80.1 &       &       &       &       &       &  78.2\\ 
5848.76 &       &       &       &  11.9 &       &       &       &       &  12.1 & \\ 
5854.62 &   6.5 &   9.2 &       &       &  11.6 &   9.3 &       &       &       & \\
6032.64 &  13.0 &  13.8 &       &       &       &       &       &       &       & \\  
6036.55 &   6.5 &   8.1 &       &       &  10.6 &   8.2 &       &       &       & \\
6048.53 &   6.6 &       &  25.2 &       &       &       &       &       &  18.3 &  22.5\\ 
6054.36 &   5.3 &       &       &       &   7.7 &   5.3 &       &       &       & \\
\hline
\end{tabular}\\
\begin{flushleft}
$^a$ The mean wavelength is given.\\
$^b$ Corrected for the contribution of \ion{Al}{iii} $\lambda$4479.93 estimated from spectral synthesis.\\
\end{flushleft}
\end{table*}  

\end{document}